\documentclass[useAMS,usenatbib]{aa}
\usepackage{txfonts}
\usepackage{natbib}
\bibpunct{(}{)}{;}{a}{}{,}
\usepackage{longtable}
\usepackage[dvips]{graphicx}
\begin{document}
\title{A study of the population of LMXBs in the bulge of M31}
\author{R. Voss\inst{1} \and M. Gilfanov\inst{1,2}}
\institute{Max Planck Institut f\"ur Astrophysik, 
Karl-Schwarzschild-Str. 1, 85740 Garching bei M\"unchen, Germany \and
Space Research Institute, Russian Academy of Sciences, Profsoyuznaya
84/32, 117997 Moscow, Russia
\\
email:[voss;gilfanov]@mpa-garching.mpg.de}
\titlerunning{LF of X-ray point sources in the bulge of M31}
\date{Received ... / Accepted ...}

\offprints{R. Voss}

\abstract{}
{
We explore the population of X-ray point
sources in the bulge of M31 to contrast
properties of various subpopulations, such as persistent and
transient sources and primordial LMXBs and dynamically formed ones.  
}
{Based on the data from 26 archival \textit{Chandra} observations we
study the source content and properties of various subpopulations of
X-ray sources  to a maximum distance of 12{\arcmin} from the centre of
M31.  
}
{
To a limiting luminosity of $\sim 10^{35}$ erg s$^{-1}$ we find
263 X-ray point sources, with $\sim 1/3$ of these being background
galaxies. 
A study of the spatial distribution and the
luminosity function of the X-ray sources shows that the distribution
of primordial LMXBs is consistent with the distribution of the 
\textit{K}-band light and that their luminosity function flattens
below $\sim 10^{37}$ erg s$^{-1}$ to the $dN/dL\propto L^{-1}$ law 
in agreement with the behaviour found earlier for LMXBs in the Milky
Way and in Cen A. Within a radius of $12\arcmin$, the
luminosity function  is independent of distance to the centre
of M31, in contrast to earlier \textit{Chandra} studies.
The LMXBs located in globular clusters and within $\sim 1\arcmin$ from
the centre of M31 are presumably created via dynamical
interactions. The dynamical origin of the $r< 1\arcmin$ sources is
strongly suggested by their radial distribution which follows the
$\rho^2_\textit{*}$ profile 
rather than the K-band light distribution.    
Their luminosity function shows a  prominent
fall-off below $\log(L_X)\la 36.5$. Although the statistics 
is insufficient to claim a genuine low-luminosity cut-off in the luminosity
function, 
the best fit powerlaw with a slope of -0.6$\pm$0.2 is
significantly flatter than the $dN/dL\propto L^{-1}$ law. 
We also searched  for transients  and found 28 sources
that varied by a factor larger than 20. Their spatial distribution
follows the distribution of the persistent LMXBs within the accuracy
allowed by the limited number of transients. 
}
{}

\keywords{
galaxies: individual: M31 -- X-rays: binaries -- X-rays: galaxies 
}

\maketitle

\section{introduction}
With the advent of \textit{Chandra}, 
X-ray point sources in nearby galaxies became a 
subject of intense study
\citep[see][and references therein]{fabbiano}.
In spiral and starburst galaxies the X-ray luminosity function
(LF) has been shown to be a powerlaw with a differential slope of $\sim$1.6
\citep{grimm}, whereas the LF in elliptical galaxies seems to have
a more complicated shape, being steep at the bright end, $\log(L_X)>
37.5$, with power law index in the $\sim$ 1.8-2.5 range, and flat below 
$\sim$ $\log(L_X)<37.0$ \citep{gilfanov,voss}. There is, however,
currently no consensus on the existence and position of the breaks, and
on the slope below a few times $10^{37}$ erg s$^{-1}$ \citep{kim}.\\

While M31 is not an elliptical galaxy, the population of X-ray sources
in the bulge mainly consists of low mass X-ray binaries (LMXBs), similar
to the population of X-ray sources in elliptical galaxies. It is
therefore fair to expect the X-ray LF to be similar to that of ellipticals
(and perhaps even more interesting if differences show up).
The LF can only be studied to a limiting luminosity of $10^{37}$ erg s$^{-1}$
in the closest large ellipticals \citep{kim}, except for Cen A
\citep{voss} that is both an unusual elliptical and suffers from contamination
of X-rays from other sources than binaries. In contrast to this, the 
proximity of M31 makes it possible to study the LF down 
to $10^{35}$ erg s$^{-1}$.\\

M31 has been observed extensively both by \textit{XMM-Newton} and 
\textit{Chandra}, and
the point source population has been analysed in a number of papers,
see e.g. \citet{kong,kaaret,pietsch2,williams,williams3}. 
In the central parts of M31 the point spread function
(PSF) of \textit{XMM-Newton} 
causes source confusion and therefore only \textit{Chandra}
observations are suited for studies of the weak sources in this region. 
The LF of the inner region of M31 has previously been studied 
with \textit{Chandra} by \citet{kong,kong2,williams}.
There are several good reasons
to reinvestigate the LF in this region. The inner bulge of M31
has been observed a number of times after this study, significantly
increasing the exposure, and also the previous study did not include the
effects of incompleteness and contamination by background sources,
which influences their conclusions significantly.\\ 

The paper is structured as follows. In section \ref{sect:data} we describe
the data sets and the basic data preparation and analysis. The
identification of sources in other wavelengths is discussed in this
section. The properties of the population of X-ray
binaries in the bulge of M31 are analysed in section \ref{sect:pop},
including the spatial distribution and analysis of incompleteness
effects. The search for and analysis of transient sources is presented
in section \ref{sect:trans}. The LFs of the source populations
are analysed in section \ref{sect:LF}. We conclude
in section \ref{sect:conc}.


\begin{table*}
\begin{center}
\caption{The specifications of the \textit{Chandra} observations used in this
  paper. The corrections given in the last two columns are the corrections
  applied to the aspect files to align the observations and} to achieve 
absolute astrometry. 1 pixel
equals 0.492{\arcsec}.
\label{obs}
\begin{tabular}{lcccccccc}
\hline\hline
Obs-ID & Date & Instrument & Exp. Time  & R.A. & Dec. & Data Mode & Correction
West & Correction North\\
\hline
0303 & 1999 Oct 13 & ACIS-I & 12.01 ks & 00 42 44.4 & $+$41 16 08.30 & FAINT  & $+$0.72 pixel & $-$0.29 pixel\\
0305 & 1999 Dec 11 & ACIS-I & 04.18 ks & 00 42 44.4 & $+$41 16 08.30 & FAINT  & $-$0.59 pixel & $-$0.16 pixel\\
0306 & 1999 Dec 27 & ACIS-I & 04.18 ks & 00 42 44.4 & $+$41 16 08.30 & FAINT  & $-$0.51pixel & $-$0.01 pixel\\
0307 & 2000 Jan 29 & ACIS-I & 04.17 ks & 00 42 44.4 & $+$41 16 08.30 & FAINT  & $-$0.34 pixel & $+$0.19 pixel\\
0308 & 2000 Feb 16 & ACIS-I & 04.06 ks & 00 42 44.4 & $+$41 16 08.30 & FAINT  & $+$0.77 pixel & $+$1.34 pixel\\
0309 & 2000 Jun 01 & ACIS-S & 05.16 ks & 00 42 44.4 & $+$41 16 08.30 & FAINT  & $-$0.41 pixel & $+$0.12 pixel\\
0310 & 2000 Jul 02 & ACIS-S & 05.14 ks & 00 42 44.4 & $+$41 16 08.30 & FAINT  & $-$0.40 pixel & $+$0.17 pixel\\
0311 & 2000 Jul 29 & ACIS-I & 04.96 ks & 00 42 44.4 & $+$41 16 08.30 & FAINT  & $-$1.29 pixel & $-$2.79 pixel\\
0312 & 2000 Aug 27 & ACIS-I & 04.73 ks & 00 42 44.4 & $+$41 16 08.30 & FAINT  & $-$0.90 pixel & $+$1.62 pixel\\
1575 & 2001 Oct 05 & ACIS-S & 38.15 ks & 00 42 44.4 & $+$41 16 08.30 & FAINT  & $-$0.97 pixel & $+$0.04 pixel\\
1577 & 2001 Aug 31 & ACIS-I & 04.98 ks & 00 43 08.5 & $+$41 18 20.00 & FAINT  & $-$2.79 pixel & $-$2.71 pixel\\
1581 & 2000 Dec 13 & ACIS-I & 04.46 ks & 00 42 44.4 & $+$41 16 08.30 & FAINT  & $-$0.62 pixel & $+$2.93 pixel\\
1582 & 2001 Feb 18 & ACIS-I & 04.36 ks & 00 42 44.4 & $+$41 16 08.30 & FAINT  & $+$1.65 pixel & $+$2.49 pixel\\
1583 & 2001 Jun 10 & ACIS-I & 05.00 ks & 00 42 44.4 & $+$41 16 08.30 & FAINT  & $-$0.50 pixel & $-$3.95 pixel\\
1585 & 2001 Nov 19 & ACIS-I & 04.95 ks & 00 43 05.6 & $+$41 17 03.30 & FAINT  & $-$1.28 pixel & $-$0.81 pixel\\
1854 & 2001 Jan 13 & ACIS-S & 04.75 ks & 00 42 40.8 & $+$41 15 54.00 & FAINT  & $-$0.74 pixel & $+$0.19 pixel\\
2895 & 2001 Dec 07 & ACIS-I & 04.94 ks & 00 43 08.5 & $+$41 18 20.00 & FAINT  & $-$0.49 pixel & $+$0.22 pixel\\
2896 & 2002 Feb 06 & ACIS-I & 04.97 ks & 00 43 05.5 & $+$41 17 03.30 & FAINT  & $-$0.20 pixel & $+$0.83 pixel\\
2897 & 2002 Jan 08 & ACIS-I & 04.97 ks & 00 43 09.8 & $+$41 19 00.72 & FAINT  & $-$0.44 pixel & $+$0.01 pixel\\
2898 & 2002 Jun 02 & ACIS-I & 04.96 ks & 00 43 09.8 & $+$41 19 00.72 & FAINT  & $-$0.56 pixel & $-$0.02 pixel\\
4360 & 2002 Aug 11 & ACIS-I & 04.97 ks & 00 42 44.4 & $+$41 16 08.90 & FAINT  & $-$0.19 pixel & $-$0.07 pixel\\
4678 & 2003 Nov 09 & ACIS-I & 04.87 ks & 00 42 44.4 & $+$41 16 08.90 & FAINT  & $+$0.06 pixel & $-$0.35 pixel\\
4679 & 2003 Nov 26 & ACIS-I & 04.77 ks & 00 42 44.4 & $+$41 16 08.90 & FAINT  & $+$0.00 pixel & $-$0.96 pixel\\
4680 & 2003 Dec 27 & ACIS-I & 05.24 ks & 00 42 44.4 & $+$41 16 08.90 & FAINT  & $-$0.36 pixel & $-$0.86 pixel\\
4681 & 2004 Jan 31 & ACIS-I & 05.13 ks & 00 42 44.4 & $+$41 16 08.90 & FAINT  & $-$0.59 pixel & $-$1.12 pixel\\
4682 & 2004 May 23 & ACIS-I & 04.93 ks & 00 42 44.4 & $+$41 16 08.90 & FAINT  & $-$0.64 pixel & $-$0.07 pixel\\

\hline
\end{tabular}
\end{center}
\end{table*}

\section{Data analysis}
\label{sect:data}

The analysis in this paper is based on 26 \textit{Chandra} ACIS
observations, with aimpoints within 10{\arcmin} from the centre of
M31 (RA~00~42~44.31, Dec~+41~16~09.4).
Information about the observations is listed in Table \ref{obs}.
The data preparation was done following the standard {CIAO\footnote{http://cxc.harvard.edu/ciao/} threads
(CIAO version 3.2.1; CALDB version 3.0.3), and limiting the energy range to 0.5-8.0 keV. 
The ACIS chips sometimes experience flares of enhanced background.
For point source detection and luminosity estimation it is not
necessary to filter out the flares, since the increased exposure
time outweighs the increased background.

We used CIAO {\tt wavdetect} to detect sources. The input parameters for
the detection procedure are similar to those used in \citet{voss}.
We detected sources within 10{\arcmin} of the aimpoint in each of
the individual observations. Due to limitations of the absolute
astrometry of \textit{Chandra}, the observations have to be aligned
before they are combined.
We chose to align the observations to OBS-ID 1575, as this is the
observation with highest exposure time. For each of the observations
we determined the number of sources matching sources found in OBS-ID
1575, excluding all ambiguous matches, such as a source in one of
the source lists being close to two sources in the other list.
The source lists were shifted relative to each other, and the smallest
rms-distance between the sources were found.
This method made it possible to align the 25 observations with OBS-ID
1575 with a minimum of 40 matches.
The corrections applied to the
observations are listed in Table \ref{obs}.
All the observations were then reprojected
into the coordinate systems of OBS-ID 1575 using
CIAO {\tt reproject\_events}, and a merged observations were created.
Notice that the steps above were taken in order to align and combine
the observations, not to achieve good absolute astrometry. This will
be dealt with using optical data in section \ref{absolute}.

An exposure map was created for each of the observations, assuming the
energy distribution to be a powerlaw with photon index of 1.7 and
Galactic absorption of 6.68$\times 10^{20}$cm$^{-2}$.
\citep{dickey}. In the following we use the same spectrum to convert the
observed count rates to unabsorbed source flux. To calculate the
luminosity of the sources, we assumed a distance of 780 kpc
to M31 \citep{dist1,dist2}.

To estimate the source counts we applied circular aperture photometry.
The output count rate for each detected source is calculated 
inside a circular region centered on the source central coordinates given 
by {\tt wavdetect}.
The radius of the circle was determined individually for each source 
so that the encircled energy
was 85\%. To find this radius we extracted the PSF using CIAO {\tt psfextract} 
task
for each of the 26 observations, and the PSFs were combined
using the values of the exposure maps as weights.
For the background region we used a circle with radius 3 times the radius
of the source region and the source region excluded, as well as the
source regions of eventual nearby sources. The corrected source counts and
errors were then found by the equations \citep{harnden}:
\begin{equation}
\label{eq:lum}
S=\frac{C(b-d)d^{-1}-Q}{\alpha bd^{-1}-\beta}
\end{equation}
and
\begin{equation}
\sigma_S^2=\frac{\sigma_C^2(b-d)^2d^{-2}+\sigma_Q^2}{(\alpha bd^{-1}-\beta )^2}.
\end{equation}
Here $S$ is the total number of counts from the source, C is the number of counts
inside the source region and Q is the number of counts in the background region,
$\alpha$ is the integral of the PSF over the source region, $\beta$ is the
integral of the PSF over the source and background regions, b is the area of
the source and background regions and d is the area of the source region.
For close sources the extraction regions can overlap. In this case a
second iteration was performed. In this iteration the number of counts
of the neighbouring sources, together with their PSF was used to find
the contamination from nearby sources of the source and background regions of a
source. This contamination from nearby sources was then subtracted from $C$ and $Q$, and
equation \ref{eq:lum} was repeated. In all cases the effect of the
contamination was small enough to justify the use of this method with
only on iteration.

\subsection{2MASS LGA}
\label{sect:2MASS}
To compare the spatial distribution of the point sources with the
distribution of mass in M31, we used a \textit{K}-band image of
the region from the 2MASS Large Galaxy Atlas \citep{2MASS}. The image
has a resolution of 1{\arcsec}, and it is therefore possible to
clearly see point sources. While many of the sources are objects inherent
to M31, such as globular clusters, some of the sources may be
foreground or background objects. In order to remove these objects, we
correlate the image with the Revised Bologna catalogue of M31 globular 
clusters \citep{bologna},
and thereby find the maximum \textit{K}-band luminosity of a globular
cluster, $\sim$ 2$\cdot 10^6 L_{K,\odot}$. All point sources more luminous than this are removed from
the image (replaced with local background). This corresponds to 14
per cent of the luminosity in the outer part of the image (the annulus
9{\arcmin}-12{\arcmin}). Point sources
with luminosities between $\sim$ 4$\cdot 10^5 L_{K,\odot}$ and $\sim$ 2$\cdot 
10^6 L_{K,\odot}$ corresponds only to 2 per cent of the total luminosity, and
most of these sources are GCs in M31. We therefore conclude that after our
removal of the most luminous sources, contamination from point sources
not in M31 is insignificant in the outer region. In the rest of the image
the luminosity density of M31 is higher and contamination is therefore
even less important.  

\subsection{Absolute astrometry}
\label{absolute}
We used the 2MASS All-Sky Point Source Catalog \citep{2MASS2} to 
achieve better absolute astrometry. This catalog was chosen, due to
the high number of (true) matches with our source list. The
astrometric precision ranges from $\sim$ 0.1{\arcsec} for brighter sources
to 0.4{\arcsec} for the weakest sources.
The X-ray sources were correlated with the 3132 2MASS point sources in
the observed region and the
X-ray image was shifted to give the smallest rms-distance for matches with
a distance less than 1{\arcsec}. This gives a correction of $-$0.97 pixel
west and +0.04 pixel north with 40 matches ($<$2 random matches expected).
We note that after the corrections, our source coordinates are in good
agreement with the coordinates given by previous studies \citep{kong,williams}.

\subsection{Source Identifications}
For the identification of the detected sources we have used a variety of
catalogues. For the identification of globular clustes we have used the
Revised Bologna Catalogue of M31 globular clusters \citep{bologna} as well
as the lists given by \citet{magnier} and \citet{fan}. GCs were divided into
the categories confirmed GCs and candidate GCs, following \citep{bologna},
and the GCs in \citet{magnier} and \citet{fan} were all considered candidates.
Planetary nebulae have been identified
using \citet{ciardullo}, and supernova remnants using \citet{magnier_snr}.
Furthermore we have searched for sources coincident with stellar novae in
\citet{pietsch}. Stars were identified using \citet{bologna} as well as
GSC 2.2 \citep{GSC} and USNO-B1 \citep{usno} (the latter two catalogues
gave only sources found in \citet{bologna} as well). Finally we searched for
possible counterparts using the NASA/IPAC Extragalactic Database (NED)
and SIMBAD. A source was assumed to be a counterpart to the X-ray
source if within a distance of 2.5{\arcsec}, except for supernova
remnants for which the distance was 5{\arcsec}.

A number of objects have been excluded from the following analysis,
4 foreground stars, 3 SNRs and one extended source (which was detected as
two sources). 
More sources have been identified with objects, as can be seen from
our source list, but the number is small, and the meaning of the
identifications is uncertain, and we have therefore chosen to keep
them in the sample. For example we note that a detailed study
of the planetary nebulae correlations \citep{williams2} suggested that
most are not true counterparts. Also of the four novae correlations,
only one (source 128) is a true counterpart (W. Pietsch, private
communication).

Our final source list consists of 263 sources
within a radius of 12{\arcmin} from the centre of M31 (Table \ref{sourcelist}).
Of these 9 sources are not included in the analysis below. Above
$10^{37}$ erg s$^{-1}$ there are 48 sources, and
above
$10^{36}$ erg s$^{-1}$ (approximately the completeness limit),
there are 136 sources included in the analysis. We expect $\sim$29
of the sources with luminosity $>10^{36}$ erg s$^{-1}$ and
$\sim$89 of all the sources to be background sources, taking into
account incompleteness, see sections \ref{CXB} and \ref{sect:incomplete}.
We find that 15 
sources are coincident with GCs (0.25 random matches expected) 
and 14 with GC candidates (1 random match expected).
\begin{table*}

\begin{center}
\caption{The list of point like X-ray sources within a
distance of 12{\arcmin} from the centre of M31. The full table
is available in the online version of the paper.}
\label{tab:sources}
\begin{tabular}{ccccrrrccccc}
\hline\hline
Number & distance & RA & DEC & cts & corrected cts & error & luminosity & type & id & name \\
(1) & (2) & (3) & (4) & (5) & (6) &  (7) & (8) & (9) & (10) & (11)\\
\hline
1
 & 1.0 & 00 42 44.37 & 41 16 08.7 & 2245 & 2580.7 & 58.0 & 7.08e+36 &  &  & r1-10\\
2
 & 2.2 & 00 42 44.38 & 41 16 07.4 & 2864 & 3308.2 & 65.4 & 1.08e+37 &  &  & r1-9\\
3
 & 4.0 & 00 42 44.38 & 41 16 05.4 & 1117 & 1233.9 & 41.3 & 5.32e+36 &  &  & r1-21\\
4
 & 4.6 & 00 42 44.30 & 41 16 14.0 & 316 & 344.9 & 22.6 & 1.52e+36 &  &  & r1-22\\
5
 & 5.4 & 00 42 43.86 & 41 16 11.1 & 225 & 244.1 & 19.2 & 1.07e+36 &  &  & r1-27\\
6
 & 7.4 & 00 42 43.87 & 41 16 03.9 & 709 & 811.6 & 33.1 & 3.53e+36 &  &  & r1-23\\
7
 & 9.7 & 00 42 44.68 & 41 16 18.2 & 906 & 1047.3 & 37.3 & 4.62e+36 &  &  & r1-8\\
8
 & 10.6 & 00 42 45.24 & 41 16 11.1 & 456 & 516.2 & 26.8 & 2.25e+36 &  &  & r1-20\\
9
 & 14.5 & 00 42 45.60 & 41 16 08.6 & 1026 & 1191.0 & 39.5 & 5.21e+36 &  &  & r1-7\\
10
 & 15.3 & 00 42 45.12 & 41 16 21.7 & 3098 & 3631.2 & 67.9 & 1.61e+37 & RAD & SIM WSTB 37W135 & r1-4\\
11
 & 20.8 & 00 42 43.88 & 41 16 29.6 & 1293 & 1463.5 & 44.5 & 6.54e+36 &  &  & r1-11\\
12
 & 21.7 & 00 42 46.01 & 41 16 19.6 & 468 & 531.3 & 27.2 & 2.33e+36 &  t &  & r1-19\\
13
 & 23.8 & 00 42 43.75 & 41 16 32.4 & 2329 & 2726.7 & 59.0 & 1.24e+37 &  &  & r1-12\\
14
 & 24.0 & 00 42 42.18 & 41 16 08.3 & 4525 & 5353.9 & 81.6 & 2.45e+37 &  t &  & r1-5\\
15
 & 25.9 & 00 42 42.48 & 41 15 53.7 & 3077 & 3601.1 & 67.6 & 1.62e+37 & PN & CIA 4 & r1-14\\
\hline
\end{tabular}\\
\end{center}
(1) -- The sequence number;
(2) -- Distance to the centre in arcsec;
(3),(4) -- Right ascension and declination of source;
(5) -- Source counts;
(6) -- Source counts after background subtraction;
(7) -- Statistical error on source counts after background subtraction;
(8) -- X-ray luminosity, 0.5-8 keV, assuming 780 kpc distance; 
(9) -- Source Type: GC -- confirmed globular cluster, GCC -- globular cluster
candidate, PN -- planetary nebula, FGS -- foreground star, NOVA -- nova,
EmO -- emission line object, RAD -- radio source, SNR -- supernova remnant,
EXT -- extended source, t -- transient source;
(10) -- precise identification and reference: Bol -- \citet{bologna}, Fan
-- \citet{fan}, Mita -- \citet{magnier}, MLA -- \citet{meyssonnier}, W2 --
\citet{williams2}, CIA -- \citet{ciardullo}, CFN -- \citet{ciardullo2}, PIE -- \citet{pietsch},
SI -- \citet{shafter}, B68 -- \citet{borngen}, SIM -- Simbad, GLG --
\citet{gelfand}, B90 -- \citet{braun}, MG -- \citet{magnier_snr}, Cra -- \citet{crampton};
(11) -- Source name in \citet{kong}, \citet{williams} and \citet{williams3};
Sources not included in these catalogues are marked with K if observed in
\citet{kaaret}, else with X, indicating that these are new sources.

\end{table*}

\section{Populations of sources in the bulge of M31}
\label{sect:pop}
\subsection{Expected numbers}
\subsubsection{Low mass X-ray binaries}
LMXBs are related to the population of old stars, and there
is therefore a correlation between their number  and the
stellar mass of a galaxy \citep{gilfanov}. 
In order to estimate the expected number and luminosity distribution
of LMXBs we used a \textit{K}-band image from 2MASS Large Galaxy Atlas
\citep{2MASS} and integrated the flux emitted in the parts of M31
analysed in this paper (excluding luminous point sources not related
to the galaxy, see Sect. \ref{sect:2MASS}). This gives a 
\textit{K}-band luminosity  
of $L_{K}=4.4\cdot 10^{10}~ L_{\odot}$. To convert it to the
stellar mass we use the color-dependent \textit{K}-band mass-to-light ratio
from \citet{bell}. For the extinction corrected optical
color of the bulge of M31, $(B-V)\approx 0.95$ \citep{Walterbos}, 
the mass-to-light
ratio is $M_{\ast}/L_K\approx 0.85$.
This gives the stellar mass of $3.8\cdot 10^{10}$ $M_{\odot}$, assuming
that the absolute \textit{K}-band magnitude of the sun is equal to
$M_{K,\odot}=3.39$.
Using the results of \citet{gilfanov} we predict $\approx$55.7 LMXBs
with $L_X>10^{37}$ erg s$^{-1}$, and $\approx$128.9 with $L_X>10^{36}$
erg s$^{-1}$.

\subsubsection{High mass X-ray binaries}
Being young objects, HMXBs are associated with star formation and, as 
expected for the bulge of a spiral galaxy, are by far a less significant
contribution to the population of X-ray binaries than LMXBs. 
Star formation is mostly associated with the disk of M31. An investigation
of the star formation rate of the disk has been conducted by
\citet{SFR_williams}, who find that the mean SFR over the last 60 Myr 
for 1.4 deg$^2$ of the M31 disk is 0.63$\pm$0.07 M$_\odot$yr$^{-1}$ 
(in the range 0.1$-$100 M$_\odot$) with
no drastic changes. Assuming a flat SFR density over the galaxy gives
an SFR of $\approx$ 0.048 M$_\odot$yr$^{-1}$ within the region analyzed
in this paper. We used the calibration of \citet{grimm} to calculate the
expected number of HMXBs  
\citep[see comment in][regarding the normalization]{shty}. 
From this we get the expectation of $\approx$0.3 HMXBs
brighter than $10^{37}$ erg s$^{-1}$, and $\approx$1.2 sources brighter
than $10^{36}$ erg s$^{-1}$.

Alternatively we have estimated upper limits for the numbers 
of HMXBs from the H$\alpha$ and FIR luminosities reported by
\citet{devereux}. For H$\alpha$ the combined luminosity from
the nuclear region and from diffuse emission inside the star
forming ring (which is at a radius of $\sim$ 50{\arcmin},
much larger than the maximum distance of 12{\arcmin} analysed in
this paper) is 4.3$\cdot$ 10$^{39}$
erg s$^{-1}$ (corrected to our distance of 780 kpc). From
\citet{grimm} we find that this corresponds to 1.1
HMXBs with a luminosity above 10$^{37}$ erg s$^{-1}$, and
4.4 HMXBs with a luminosity above 10$^{36}$ erg s$^{-1}$.
The FIR luminosity in this region is 5.25$\cdot$ 10$^{8}$ L$_\odot$,
which corresponds to 2.0
HMXBs with a luminosity above 10$^{37}$ erg s$^{-1}$, and
8.0 HMXBs with a luminosity above 10$^{36}$ erg s$^{-1}$.
It should be noted, however, that the region these luminosities
are found from is much larger than the region containing our
X-ray data, and that it is very likely that the main part of
the light is not produced by star formation, as \citet{devereux}
found that for the central region the number of O-type stars is
a factor of $\sim$ 200 lower than what would be expected if the
luminosities were due to star formation.  
We can therefore safely ignore the contribution of HMXBs in the
following analysis.

\subsubsection{Background X-ray sources}
\label{CXB}

To estimate the number of background sources, we use
results of the CXB $\log(N)-\log(S)$ determination by
\citet{moretti}. We use the source counts in the soft and
hard bands (their Eq. 2) and  convert the fluxes to the
0.5--8.0 keV band, assuming a powerlaw spectrum with a photon index of
1.4. For the total area of our survey of 0.126 deg$^2$  we 
obtain from the source counts in the soft band 
$\sim$1.8 CXB sources above the flux corresponding to 
$10^{37}$ erg s$^{-1}$, and $\sim$29 above $10^{36}$ erg s$^{-1}$.
From the hard band counts the predicted numbers are 
$\sim$1.2 and $\sim$30.5 sources. 
The predictions based on the soft and hard $\log(N)-\log(S)$ differ
slightly because of the well recognized fact that source counts in different
energy bands and flux regimes are dominated by different types of
sources, see \citet{voss}. To find the total number of background sources
in our source list, we multiply the CXB LF by the incompleteness
function found in section \ref{sect:incomplete}, and integrate over the observed luminosity range.
We find the total number of CXB sources to be 89.

\subsection{The spatial distribution of the point sources}
\label{sect:spatial}
\begin{figure}
\resizebox{\hsize}{!}{\includegraphics[angle=270]{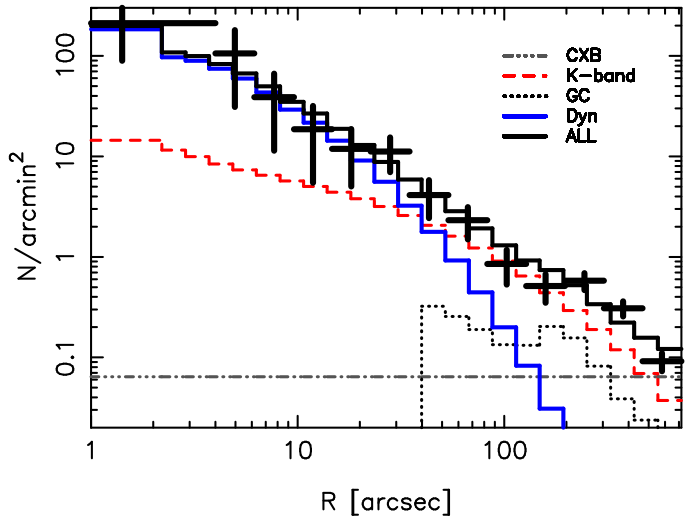}}
\caption{The spatial distribution of the point sources (crosses), 
compared
to a model consisting of primordial LMXBs, CXB sources and LMXBs in GCs.
The GC LMXB normalization was determined from the observed number of matches,
and the primordial LMXB and CXB normalizations were found from a fit to
the observed source distribution outside 1{\arcmin}. In the inner
1{\arcmin} the sources follow the expected distribution of LMXBs formed
through dynamical interactions, $\rho_{\ast}^2$ \citep{Voss2}.}
\label{fig:spatial}
\end{figure}

We studied the azimuthally averaged radial distribution
of the X-ray point sources. As the two main contributions of sources
are the LMXBs and CXBs, we model the distribution as a superposition
of two functions, representing these contributions. As the spatial
distribution of the globular clusters in M31 is significantly different 
from the mass distribution in the inner parts of the bulge, we have
accounted for the globular cluster sources separately.
The distribution
of LMXBs is assumed to follow the \textit{K}-band light, and for this the
image from the 2MASS LGA was used, whereas the density
of CXBs can be assumed to be flat on the angular scales under consideration
here. The only free parameter of the model is the ratio of normalizations
of the LMXB and CXB fraction.

The model was compared to the observations for sources more luminous
than $10^{36}$ erg s$^{-1}$ (Figure \ref{fig:spatial}), 
as sources of such luminosity could be
observed in the entire image without it being necessary to consider
incompleteness effects. This analysis was presented in \citet{Voss2},
where the data was shown to deviate significantly from the model in
the inner $r\la 1\arcmin$.
It was also shown that the discrepancy could be succesfully modelled
by binaries created through dynamical interactions.

In this paper we adopt the normalizations of the primordial LMXBS and
CXBs found in \citet{Voss2}. Above a luminosity of
10$^{36}$ erg s$^{-1}$
this corresponds to 29 CXBs and
64 primordial LMXBs, as well as $\sim$ 20 LMXBs created via dynamical
interactions in the inner bulge and $\sim$ 20 LMXBs in GCs.
The normalization of CXBs is consistent with the expectations.
From this we find that the ratio of primordial LMXBs with
luminosity above 10$^{36}$ erg s$^{-1}$ (10$^{37}$ erg s$^{-1}$) 
to stellar
mass is $N_x/M_{\ast}$ is 17.0$\pm$1.8 (8.9$\pm$1.6) sources per 
10$^{10}$ $M_{\odot}$, and the ratio of primordial LMXBs to
the \textit{K}-band luminosity is $N_x/L_{K}$ is 19.7$\pm$2.1
(10.3$\pm$1.9) sources per 10$^{10}$ $L_{\odot,K}$.
The normalization of the primordial LMXBs is about two times
smaller than the number obtained by \citet{gilfanov}. There are
two reasons for this. 1) We removed LMXBs in GCs and in the inner
bulge from the analysis, to only account for LMXBs thought to be
primordial. This was not done by \citet{gilfanov}. 2) We assumed
the \textit{K}-band mass to light ratio of the bulge of M31 to be 
0.85, as compared to the ratio of 0.56 used in \citet{gilfanov}.  

\subsection{Incompleteness}
\label{sect:incomplete}
The variations of the diffuse background level and deterioration of
the PSF at large off-axis angles lead to variations of the point-source
sensitivity across the \textit{Chandra} images. An image
in which observations with different pointings are combined has
very non-uniform exposure. As a result there are strong incompleteness
effects at the faint end of the luminosity function. We calculate the
incompleteness function for each analysed area, using the method
described in \citet{voss}, in which the incompleteness function is
calculated separately for LMXBs and for CXBs. A completeness limit
is calculated for each pixel and weighted by the expected distribution
of sources (using the same distributions as in section \ref{sect:spatial}).
In Figure \ref{fig:incomp} the incompleteness function is shown for 
both LMXBs and CXBs in 3 regions. As in \citet{voss} we verified
our calculated incompleteness functions by simulations of the type
used by \citet{kim2}. In each of the simulations we used Monte Carlo
techniques to simulate 10\,000 point sources. Each of the sources
was placed on the real image of M31, according to the expected spatial
distribution of the source type (LMXBs or CXBs), and our observation
pipeline was applied to the image to test if the source is detected,
and if so, with what luminosity. This way, the observed number of
sources in a luminosity range was compared to the simulated
number to determine the detection efficiency. For more details on the method,
see \citet{voss}.
The results of the simulations are compared
to the calculated functions for two regions in Figure \ref{fig:corsim}.  
We calculated the incompleteness function for GCs and
GC candidates by finding the detection limit at the position of
each of the GCs, and assuming the probability of containing an LMXB
to be the same for all the GCs. The incompleteness functions calculated
in this section will be used in the analysis of the LF of the LMXBs
in section \ref{sect:LF}. 

\begin{figure}
\resizebox{\hsize}{!}{\includegraphics[angle=0]{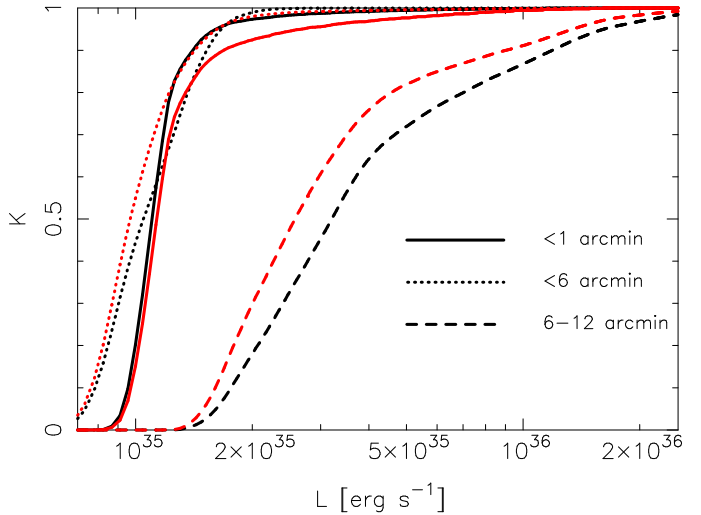}}
\caption{The incompleteness as a function of source luminosity 
for 4 regions of the bulge of
M31. The red lines show the function for LMXBs, while the black
lines show the functions for CXBs.}
\label{fig:incomp}
\end{figure}

\begin{figure}
\resizebox{\hsize}{!}{\includegraphics[angle=270]{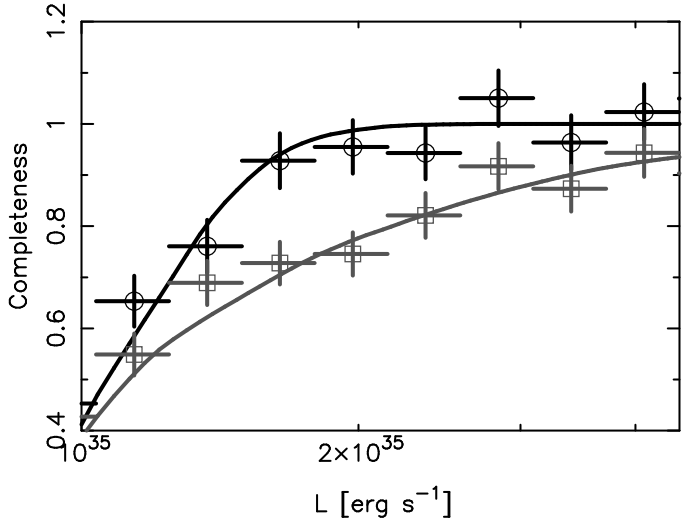}}
\caption{The results of incompleteness simulations compared to
the calculated incompleteness functions of LMXBs. The black line and
data points (circles) corresponds to the annulus with 3{\arcmin}-6{\arcmin} 
distance
from the centre of M31, while the grey line and data points (squares)
corresponds to the region within 12{\arcmin}. In both simulations
10\,000 sources were simulated.}
\label{fig:corsim}
\end{figure}

\section{Transient sources}
\label{sect:trans}
A large fraction of the sources in our sample are variable. For most
of the sources, the luminosity varies within a factor of a few.
In the combined image the luminosity is the average of the
luminosities of the single observations, weighted by the exposure.
For the sources with a low amplitude of the variability, this
weighted average is adequate for the analysis carried out in this
paper.\\

However, the study of \citet{williams3} has shown that, on average, there
is $\sim$ one transient source per observation. In an image combined from
many observations the effects of these sources on the normalization
and shape of the LF are non-negligible. As the
luminosity of a source is weighted by the exposure, it is
straightforward that the more observations that are combined, the
more transients there are, and the lower the average luminosity of each
of them will seem to be. To find and investigate transient sources,
we analysed each observation individually in the same way as the 
combined observation.

For each source it was noted for which observations it was
found and with what luminosity. If the source was not found
with {\tt wavdetect} (and if the source region had any exposure) 
we put the source into one of two categories. 
The number of photons from the source region, $N_{ps}$
(85 per cent of PSF centered on the observed coordinates of
the full image), and background region $N_{pb}$ (annulus with radius 1--3 
times the source region) were counted. 95 per cent confidence
limits $C_{low}$ and $C_{high}$ were calculated on the source 
counts \citep{gehrels}.
If $C_{low}$ was larger than the number of expected
background photons in the source region, scaled by exposure and
numbers of pixels from the background region, we labelled it
a low-significance source and used $N_{ps}$-$N_{pb}$ as source
photons to find the luminosity.
If $C_{low}$ was lower than the expectation of background
photons, we calculated an upper limit to the luminosity from
$C_{high}-N_{pb}$. Based on the considerations above we divided
the sources into three types for each observation. 
Type 0 is a source observed with
{\tt wavdetect}, type 1 is a low-significance source, and type
2 is a source not observed, for which an upper limit is given.
In Table \ref{tab:trans} the source type is given for the
observation with lowest luminosity for each source.
\\

 The ratio between the
highest and the lowest luminosity (or upper limit) was then noted for 
each of
the sources. We chose to label sources with a ratio $>$20
as transients. This gave the 28 sources listed in Table \ref{tab:trans}.
Out of these, only one was actually observed at the lowest
luminosity, indicating that the majority of these sources are
real transients. For many of our sources, the ratio limit is
set so high that if they are transients, they would not be 
labeled as such, due to lack of exposure. If the limit was
set to be lower, however, there would be a large number of
sources that are variable, but not transients, that would
be included. The amplitude of the variability
of a source is artificially enhanced by statistics when the number of
observations is large.

\begin{figure}
\resizebox{\hsize}{!}{\includegraphics[angle=270]{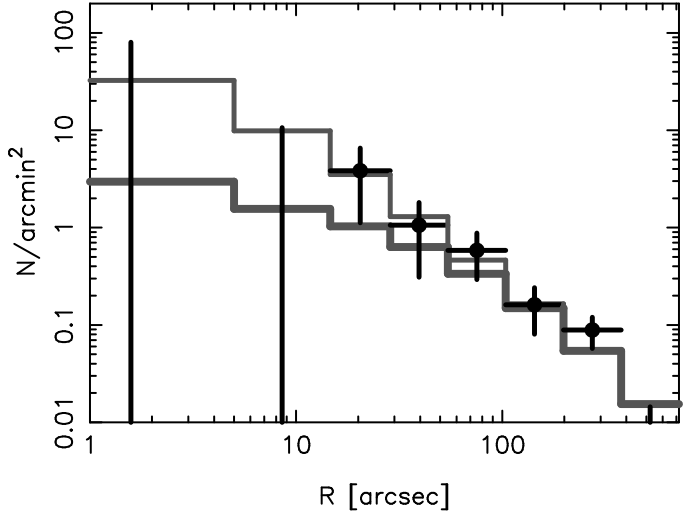}}
\caption{The spatial distribution of the transient sources, compared
to two models, the distribution of primordial LMXBs (thick grey line)
and the combined distribution of primordial LMXBs and LMXBs created
through dynamical interactions in the inner bulge (thin grey
line). The error bars are 1 $\sigma$.}
\label{fig:spatial,trans}
\end{figure}

\begin{table}
\caption{Highly variable sources with $L_{max}/L_{min}>20$. Given here are (1) the source number
from Table 1, (2) the ratio of highest observed luminosity to the lowest
observed luminosity (or upper limit), (3) the highest observed luminosity,
(4) the observation in which the luminosity was highest, (5) the observation
in which the luminosity was lowest, and (6) the observational type at
lowest luminosity, see description in text. It is also noted in this
column if the source belongs to a GC or GC candidate.}
\label{tab:trans}
\begin{tabular}{rrrrrr}
\hline\hline
Source & high/low & max lum & high obs & low obs & type\\
(1) & (2) & (3) & (4) & (5) & (6)\\
\hline
12 & 22.0 & 7.61e+36 & 1585 & 310 & 2\\
14 & 468.8 & 7.28e+37 & 308 & 1575 & 2\\
23 & 50.3 & 1.26e+37 & 4682 & 305 & 1\\
29 & 126.2 & 4.42e+37 & 1575 & 310 & 2\\
32 & 65.2 & 1.37e+37 & 4681 & 1575 & 2\\
38 & 99.5 & 2.28e+37 & 4679 & 303 & 2\\
41 & 28.3 & 3.18e+36 & 303 & 1575 & 1\\
44 & 35.7 & 3.22e+36 & 303 & 1575 & GCC,1\\
47 & 64.6 & 1.99e+37 & 4682 & 1575 & 2\\
51 & 87.1 & 1.11e+37 & 306 & 1575 & 2\\
59 & 589.0 & 4.75e+37 & 4682 & 1575 & 2\\
68 & 370.5 & 1.93e+37 & 310 & 1575 & 2\\
72 & 21.2 & 1.34e+37 & 309 & 4679 & 1\\
84 & 54.5 & 7.68e+36 & 1854 & 1575 & 2\\
85 & 79.0 & 2.56e+37 & 303 & 1585 & 1\\
90 & 58.6 & 5.36e+37 & 311 & 312 & 1\\
105 & 954.2 & 3.27e+38 & 1575 & 303 & 2\\
118 & 108.4 & 3.86e+37 & 4681 & 303 & 2\\
128 & 46.1 & 1.50e+37 & 2896 & 1575 & 2\\
130 & 104.4 & 1.49e+37 & 311 & 1575 & 2\\
136 & 92.2 & 4.55e+37 & 4682 & 305 & GCC,2\\
146 & 212.4 & 3.77e+37 & 4681 & 1575 & 2\\
155 & 96.3 & 4.48e+37 & 1585 & 1575 & GCC,0\\
212 & 65.2 & 2.68e+37 & 4682 & 1575 & 2\\
216 & 20.8 & 1.04e+37 & 1854 & 1575 & 2\\
234 & 66.9 & 1.18e+37 & 4682 & 2898 & 1\\
237 & 37.5 & 7.48e+37 & 2897 & 307 & 1\\
250 & 53.4 & 7.00e+37 & 2896 & 305 & GC,2\\
\hline
\end{tabular}
\end{table}

A catalogue of transients in M31 was published by
\citet{williams3}, and 5 further transients were found in 
a series of papers by the same group
\citep[e.g.][]{williamsT4}.
From their lists, 36 of the sources are
within the region analysed in this paper. Of these, 18
coincide with sources in our transient list. We have therefore
identified 10 new sources. Of the 18 sources remaining in
their source list, 5 of them were detected in our observations,
but did not fulfil our criteria for being transients
(they also had high/low ratios lower than 20 in
\citet{williams3}). One source (source 214 in Table \ref{tab:sources}) was labelled a transient
in their paper with a high/low ratio of 96 but only
varied by a factor of $<$10 in our observations.
The last 12 sources from their source list are not active
in our observations. \citet{trudolyubov2} detected 4 transients
with \textit{XMM-Newton}. 3 of these are not active in our
observations, while the fourth (source 234 in Table \ref{tab:sources})
was found to be transient in our observations, as well as by
\citet{williamsT4}.

We investigated the spatial distribution of the transient
sources. In Figure \ref{fig:spatial,trans} we compare their
radial distribution with two models: 1) the distribution of
the \textit{K}-band light, representing the primordial LMXBs,
and 2) the distribution of all the observed LMXBs (all sources
with CXBs subtracted). 
It can be seen that with the current set of observations, both 
models can explain the distribution. With more observations it
might be possible to distinguish between the models, and thereby
learn if the ratio of transient to persistent sources is different
for primordial LMXBs and dynamically formed LMXBs.
The
number of individual observations and length of these varies strongly from
region to region, and this can have significant effects
on the observed distribution. Also inside 5{\arcsec} the source
density is so high that transients could easily be missed.\\

\section{The luminosity function of the point sources}
\label{sect:LF}
The LF in the bulge of M31 has previously been
studied with \textit{Chandra} by \citet{kong,kong2,williams},
but for several reasons it is
interesting to do further work on this. The exposure of the
inner region has increased significantly since the previous
studies, and methods for incompleteness correction have been
developed. It is therefore possible to probe the LF at much
lower luminosities. Furthermore the previous studies neglected
the contribution of background sources. This can be important
for the outer parts of the bulge, where the density of LMXBs is
comparable to the density of background objects, see Figure 
\ref{fig:spatial}.\\

Here we study the LF of the LMXBs in detail, statistically
taking into account the CXBs, and correcting for incompleteness
as described in section \ref{sect:incomplete}.
The LMXB LFs presented
in this section are corrected by subtracting the LF of
CXBs multiplied by the incompleteness function of the CXBs.
The normalization of the LF of CXBs was chosen as in 
\ref{sect:spatial}. The LFs of the LMXBs were then corrected
by dividing by the incompleteness function of the LMXBs.
From Figure \ref{fig:incomp} it can be seen that there is
a factor of $\sim$10 difference between the sensitivity in
the regions inside 6{\arcmin} and the regions outside. At the
same time the density of X-ray sources is much higher in the
inner region than
in the outside region, making the CXB contribution less
important. In the inner regions the LF can therefore be
determined directly down to a few times $10^{35}$ erg s$^{-1}$.
We choose to present the
functions as differential LFs, as opposed to the cumulative LFs
often used in the literature. The advantage of this is that bins
are independent, and features in the LF are therefore more visible,
and easier to interpret.
The disadvantage is that it is necessary to bin the data.
The LFs presented below are cut off at a lower luminosity.
This luminosity corresponds to the limit at which the
incompleteness correction is $>2.5$ for either the
CXB LF or the LMXB LF. For this reason the LFs of the
individual regions begin at different luminosities in
figures \ref{fig:lumfun_all}-\ref{fig:lumfun_trans}.
For each of the LFs, we give the number of sources
included in the calculations. However it should be noted
that due to the corrections for incompleteness and CXB
sources applied to the LFs, the error bars in the figures
provide better estimates for the precision of the functions.
\\

\begin{figure}
\resizebox{\hsize}{!}{\includegraphics[angle=0]{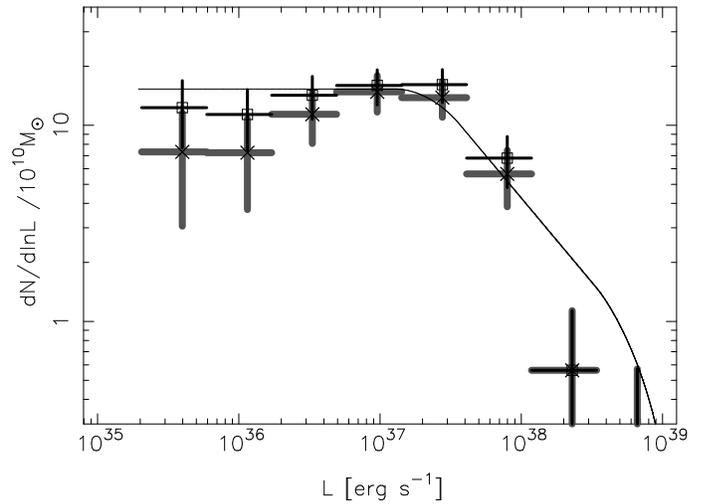}}
\caption{The LF of LMXBs within 12{\arcmin} of the centre of M31.
Squares are obtained from all the observed sources (207 sources),
whereas for the crosses, the transient sources are
excluded (179 sources). Note that in the latter case the relative contribution
of sources thought to have dynamical origin is increased, see
the discussion in the text. 
The solid line is the average LF of LMXBs in nearby galaxies from
\citet{gilfanov}. The normalization is arbitrary, but the same as
figures 5-10.}
\label{fig:lumfun_all}
\end{figure}

In Figure \ref{fig:lumfun_all} we show the LF of the entire region
within a distance of 12{\arcmin} from the centre of M31. The squares
correspond to all the sources, whereas to produce the
crosses, the transient sources (Table \ref{tab:trans}) 
were excluded. There is a clear break at $\sim 2\cdot10^{37}$
erg s$^{-1}$, consistent with previous results obtained with
\textit{Chandra} \citep{kong,kong2,williams} as well as other
X-ray telescopes \citep{primini,shirey,gilfanov}.

As it can be seen in Figure
\ref{fig:lumfun_all} the effect of 
transients is to artificially steepen the slope of the 
LF below $\sim 10^{37}$ erg s$^{-1}$. A maximum likelihood (ML)
fit by a single powerlaw in the range 
$2\cdot 10^{35}- 10^{37}$ erg s$^{-1}$ gives a slope
of $-0.85\pm 0.12$ and $-0.7\pm 0.16$ with and without transients,
respectively. 
While the difference is not statistically significant in our
sample, it is a
systematic effect that should not be ignored in general when studying
a large number of observations combined.
To avoid distortion of the LF  due to transient sources, they
should be excluded from the 
analysis, and we have done so in the rest of the analyses presented in
this paper.

After the transient sources have been excluded, the faint
end ($2\cdot10^{35}-10^{37}$ erg s$^{-1}$) of the LF appears to be
significantly flatter than $L^{-1}$ powerlaw.
This is caused by the sources located in globular clusters and in the
inner bulge, thought to have dynamical origin, as the LF of these sources
has a prominent fall-off at low luminosities (Figure
\ref{fig:lumfun_dyn}). Their relative contribution to 
the LF increases when the transient sources are excluded.
As will be shown below, the LF of the primordial sources is 
consistent with the general shape found by \citet{gilfanov}.

In Figure \ref{fig:lumfun_dyn} the LFs of the LMXBs thought to be
created through dynamical interactions are presented. The
LMXBs from the inner 1{\arcmin} of the M31 bulge are shown (crosses), 
compared to the LFs of LMXBs in confirmed GCs (squares) and
GC candidates (triangles). The three LFs are consistent with
each other, and
for all three populations the number of LMXBs falls off at 
luminosities below
$\log L_x \lesssim 36.0-36.5$. This is most significant
in the bulge population, which can also be observed to the lowest
luminosity level. For the GC candidate sources, the falling off
at low luminosities is hardly significant, but it is known that
the GC candidate list is contaminated by background galaxies \citep{bologna},
and with the LF of CXB sources, the effect of such a contamination
would be to raise the lower end of the LF.\\

In the inner 1{\arcmin} of M31, as well as in the GCs, the source density is
so high that source blending can become a factor.
We performed Monte Carlo simulations of the source population in 
the inner 1{\arcmin} of
M31, similar to the ones performed to estimate incompleteness (see section
\ref{sect:incomplete}). We assumed the average luminosity function of
\citet{gilfanov}, with the normalization according to our observed number
of sources in the region, and with a lower cut-off at $10^{36}$ erg s$^{-1}$
as observed, and the spatial distribution of all sources in Figure 
\ref{fig:spatial}. From this we find that the fraction of blended sources
(parameter $b$ in appendix \ref{sec:blending}) is $\sim 3-4$\%. For an
alternative luminosity function in which the lower cut-off is set at
$10^{34}$ erg s$^{-1}$, $b\sim$9-10\%.\\
As only $\sim$20\% of the GCs host LMXBs, the
fraction of blended sources is also likely to be low here ($b \sim 4$ per
cent, assuming that all GCs are identical, but the exact number
depends on the distribution of GC properties
relevant for the formation of LMXBs). For comparison one out of 12 GCs hosting
LMXBs in the Galaxy has been shown to host two LMXBs \citep{white2},
corresponding to $b\sim 8-9$ per cent. In
appendix \ref{sec:blending} we consider the effects of source blending on
the observed LF, and we show that for the values of $b$ in this range,
the effect of blending is not important.\\
Given the \textit{Chandra} angular resolution, at the distance of M31, 
all X-ray sources in a GC will be blended into
one point-like source. As there are numerous sources of low luminosity 
$L_X\lesssim10^{34}$ erg s$^{-1}$, this could affect our
analysis. This is not the case, however, as the luminosities of these
sources are too low. For example the combined luminosity of the $\sim 300$
observed sources in the massive Galactic GC 47 Tucanae is $\sim5\cdot10^{33}$
erg s$^{-1}$ \citep{heinke}, i.e. less than 1\% of the luminosity of
a typical GC source observed in M31.\\

From a comparison of Figure \ref{fig:lumfun_all} and Figure
\ref{fig:lumfun_dyn}, it appears that the LF of the LMXBs of presumably
dynamical origin is different from the average LF of all
the LMXBs.
We investigate this difference further
in Figure \ref{fig:lumfun_almost} where
we compare the LF of the
dynamically formed LMXBs (sources located in the inner 1{\arcmin} and
in confirmed GCs) with the LF of all other sources 
in the 1{\arcmin}-9{\arcmin} annulus. These are, presumably, of primordial origin. 
This figure confirms qualitatively the difference between the two
populations. Due to rather limited numbers
of sources the LFs are not very tightly constrained. In particular,
the statistics is insufficient to discriminate between a genuine
low-luminosity cut-off in the LF of dynamically formed sources and its
moderate flattening.
To estimate the statistical significance  we consider 
the numbers of sources in different sub-populations in the
$1.5\cdot10^{35} - 10^{36}$ erg s$^{-1}$ 
luminosity range (the lower boundary is defined by the low bound of the
primordial LF, see discussion earlier in this section).
There are two sources in this luminosity range in the population of the
dynamically formed LMXBs, whereas $17.4\pm4$ would be expected if the
source counts in the $10^{36} - 10^{37}$ erg s$^{-1}$ range were
extrapolated with a $dN/dL\propto L^{-1}$ law. Due to nearly identical
normalizations of the two LFs above $\log(L_X)\ga 36$ (cf. Figure
\ref{fig:lumfun_almost}), these numbers can be directly 
compared with $32\pm11$  primordial sources (CXB contribution 
subtracted and incompleteness corrected) observed in the same
luminosity range ($20\pm5.6$ sources expected for $L^{-1}$
extrapolation). 
In order to further quantify the difference between the two luminosity
distributions we fit them with a single powerlaw
in the 1.5$\cdot10^{35} - 10^{37}$ erg s$^{-1}$ luminosity range, using
ML fits. For the primordial sources we obtained a differential
slope of $-1.11\pm 0.18$, while the LF of dynamicaly formed LMXBs  has a
slope of $-0.6\pm$ 0.2. Although the difference between these two
numbers is only marginally significant, the LF slope of the dynamically
formed LMXBs is  inconsistent with the value of $-1$ obtained for the
average LMXB LF. 

\begin{figure}
\resizebox{\hsize}{!}{\includegraphics[angle=0]{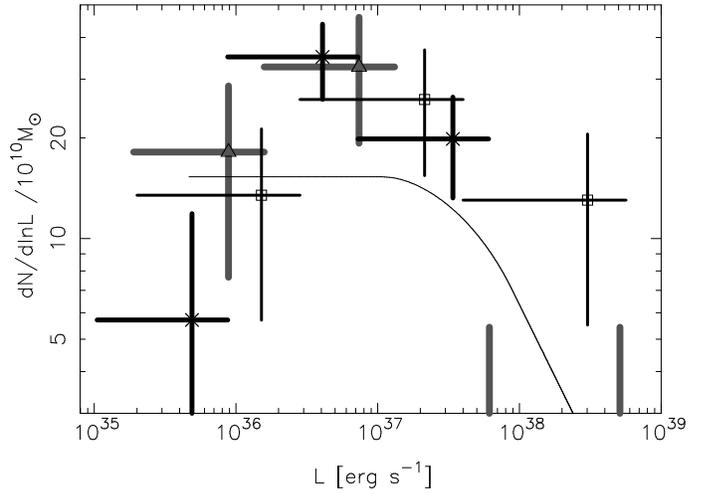}}
\caption{The LF of the sources in the inner 1{\arcmin} (crosses, 27 sources),
the LF of the X-ray sources in confirmed GCs (squares, 12 sources) and 
the LF of
the X-ray sources in GC candidates (triangles, 9 sources). 
The normalization
of the LFs from GCs and GC candidates is arbitrary.
The solid line is the average LMXB LF, with the same normalization in
figures 5-10.}
\label{fig:lumfun_dyn}
\end{figure}

It is interesting to compare the results with the bright ($L_x\gtrsim10^{35}$
erg s$^{-1}$) LMXBs in Galactic
GCs. Currently 13 of such LMXBs have been observed in 12 GCs, and due to the
proximity the sample is believed to be complete, except for possible
future transients. While all of these sources have been observed with
\textit{Chandra}, there are not published luminosities for all of them,
and analysis of the observations are beyond the scope of this paper.
Instead we find the luminosities by averaging the lightcurves for each
of the sources from \textit{RXTE ASM}, over all of the observed time
(until January 1st, 2007). The count rates were converted to fluxes
in the 0.5-8.0 keV band, assuming a powerlaw spectrum with photon
index 1.7, using 
{\tt PIMMS}\footnote{http://cxc.harvard.edu/toolkit/pimms.jsp}.
This gives a conversion factor of 
1~count~s$^{-1}=4.3\cdot10^{-10}$~erg~cm$^{-2}$~s$^{-1}$.
In Figure \ref{fig:lumfun_MW} the LF of the LMXBs in Galactic GCs is
compared to the LF of LMXBs in confirmed GCs in M31, and it is shown
that in the Galaxy there is a clear cut-off at $\sim 10^{36}$ erg 
s$^{-1}$.
\\

\begin{figure}
\resizebox{\hsize}{!}{\includegraphics[angle=0]{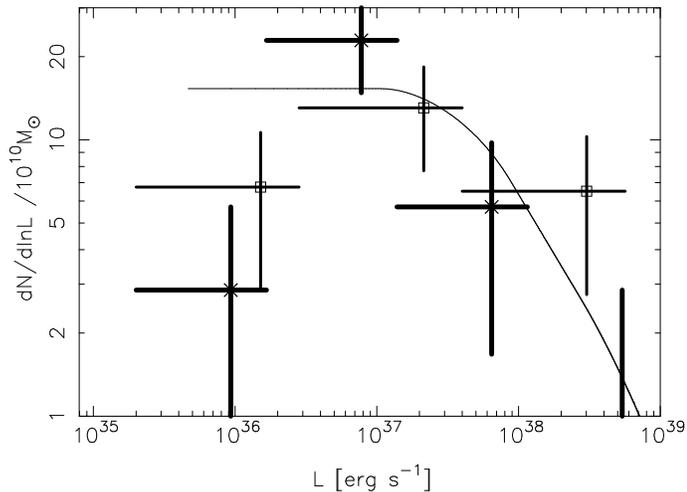}}
\caption{The LF of the LMXBs in Galactic GCs (crosses, 12 sources) compared
to the LF of the LMXBs in confirmed GCs in M31 (squares, 12 sources). 
The solid line
is the average LMXB LF, with the same normalization in figures 5-10.}
\label{fig:lumfun_MW}
\end{figure}

The difference between the LF of primordial and dynamically formed
LMXBs is interesting for several reasons. 
It has recently been discussed whether most of the field LMXBs
were actually formed in GCs \citep[e.g.][]{white,juett}. 
If the LFs of the field  and GC LMXBs are different, it is a
strong indication that their origin is different as well.
Moreover, the fact that the LF of the sources in the
inner 1{\arcmin} is consistent with the LF of the GC sources and
different from the LF of the field LMXBs reinforces the
conclusion of \citet{Voss2} that the surplus X-ray sources
in the center of M31 are LMXBs created through dynamical interactions
in high stellar density environment of the inner bulge.

Differences between the LFs of dynamically formed LMXBs in the
globular clusters and field LMXBs have previously been reported
\citep{kong,trudolyubov} but at larger luminosities, above $10^{37}$
erg s$^{-1}$. These have been disputed in a recent thorough study of six
elliptical galaxies by \citet{kimE}, who concluded that the LFs
of the two populations are consistent. We note that the differences found
in the analysis of this paper occur at luminosities below the sensitivity
threshold of \citet{kimE}, and our results therefore do not
contradict theirs.\\

Various models for LMXB evolution exist, from which the
shape of their luminosity distribution can be predicted. For a
population of LMXBs with non-degenerate 
donors, the differential powerlaw slope of $-1$
at luminosities below $\sim 10^{37}$
can be naturally obtained if the
mass transfer is driven by gravitational radiation, as
opposed to the steeper slope above $\sim 10^{37}$ erg s$^{-1}$,
which can be explained by the magnetic braking driven systems
\citep{postnov,pfahl}. 
Ultra-compact X-ray binaries (UCXB) have degenerate donor stars and the
mass transfer is driven by gravitational radiation 
alone. In this case the reaction of the WD donor to mass
loss is important for the mass transfer rates and therefore also for
slope of the LF, and models have been succesful in explaining
the bright end of the LMXB LF, near and above $\sim 10^{38}$ erg s$^{-1}$ 
\citep{bildsten}. 
No modeling of the fainter end of luminosity distribution for
the UCXB population has been reported so far. Intuitively, one might
 expect that the luminosity distribution of these systems
should fall off at low luminosities. Although the UCXB systems are
very unlikely to contribute significantly to the bulk of fainter
primordial LMXBs in the $\log(L_X)\la 37$ luminosity domain, their
importance increases dramatically in the entire luminosity range when
considering the LMXBs of dynamical origin, especially those formed in
the high velocity environment of the inner bulge \citep{Voss2}.
This offers a plausible explanation for the rather peculiar shape of the
luminosity distribution of the globular cluster sources and of the
sources in the inner $1\arcmin$ of M31.  
As the reaction of the WD donor to mass loss 
depends on the chemical composition of the WD, modeling the
luminosity function at low luminosities and comparing with
observations of LMXB in the inner bulge and in
globular clusters in M31 and other galaxies might reveal new
information on the progenitors of the UCXBs and advance our
understanding of binary evolution and dynamical interactions in dense
stellar environments in general. 

Another factor, potentially important at low mass transfer rates, is the
thermal-viscous instability, which causes transient behaviour in LMXBs
below some critical value of the mass accretion rate
\citep{paradijs}. Consequently, the LF of persistent sources should be
expected to have  a break around this luminosity. The critical 
luminosity is somewhere in the  $\la 10^{35-36}$ erg s$^{-1}$ domain,
and depends, among other parameters, on the physical size of the
accretion disk around the compact object \citep{king}.
In this picture, if the disk instability was the reason for the
observed low luminosity cut-off observed in Figure
\ref{fig:lumfun_dyn}, the critical luminosity for UCXBs would be  
expected to be lower than for LMXBs with non-degenerate donors. 
This prediction seems 
to be in contrast to the result of this paper, that the LF of the
dynamically formed LMXBs (presumably having a significantly  higher 
fraction of UCXBs) appears to be flatter (i.e. fewer faint
systems) than the LF of the primordial LMXBs.\\

In the previous study of LMXBs in the bulge of M31
\citep{kong}, it was found that their LF varied significantly with
the distance from the centre, becoming progressively steeper with
radius. We searched for the radial trend by
comparing the LFs of the primordial LMXBs (that is, with LMXBs 
in GCs excluded) of the annuli 
1{\arcmin}-3{\arcmin}, 3{\arcmin}-6{\arcmin} and 6{\arcmin}-12{\arcmin}
(Figure \ref{fig:lumfun_regs}), 
and found no statistically significant variations.
We suggest that the difference 
in the LF reported by them,
especially between their regions 2 and 3, 
was caused by the contribution of CXB sources which becomes more
larger in the outer parts of the bulge 
(cf. Figure \ref{fig:spatial}). Note that this possibility was
also considered by \citet{kong}.
\\

In Figure \ref{fig:lumfun_trans}, we show the LF of  transient
sources. As the average luminosity is meaningless, due to its
dependence on the exposure time and pattern of the observations in which
the sources were found, we have chosen to use the maximum of
the observed luminosities in the individual observations 
for each source. The LF of
these sources follows the average LF of LMXBs to a
minimum luminosity of 10$^{37}$ erg s$^{-1}$. Below this
the observed distribution falls off quickly, but this
is likely an artifact of the our selection criterion for
transient sources, $F_{max}/F_{min}>20$. 
For most sources with peak luminosities below 10$^{37}$ erg s$^{-1}$,
it is not possible to constrain the quiescent luminosity well
enough to classify the sources as transients.

\begin{figure}
\resizebox{\hsize}{!}{\includegraphics[angle=0]{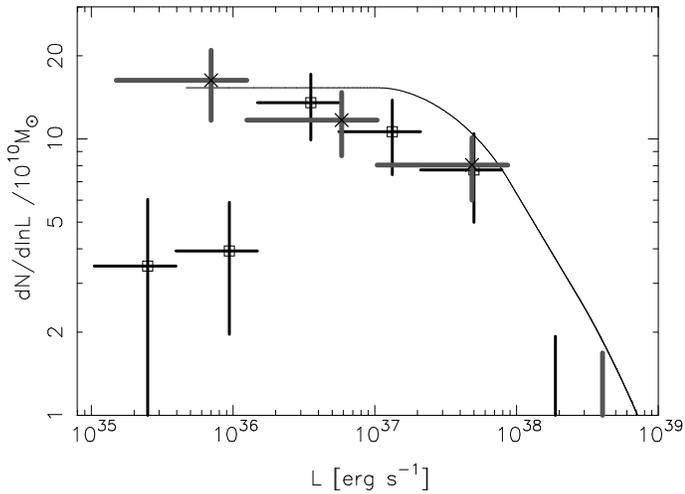}}
\caption{The LF of sources with a radial distance of 
1{\arcmin}-9{\arcmin} from the centre of M31, excluding sources in
GCs and GC candidates (crosses, 130 sources),
compared to the LF of sources from the inner 1{\arcmin}
and sources in confirmed GCs (squares, 40 sources). 
The normalization of the
latter is arbitrary. 
The solid line is the average LMXB LF, with the same normalization
in figures 5-10.}
\label{fig:lumfun_almost}
\end{figure}

\section{Conclusions}
\label{sect:conc}
We have studied the X-ray point sources in the bulge of
M31 ($r<12${\arcmin}) based on archival \textit{Chandra} data.  
Our study concentrated
on statistical properties of the population, with emphasis on the
spatial distribution and LFs of the various subpopulations. One of our
primary goals was to contrast properties of different sub-populations
of X-ray point sources, namely transient and persistent sources,
primordial LMXBs and dynamically formed ones.
To achieve this we combined  26 \textit{Chandra} observations to
obtain as much 
exposure as was possible and
implemented an adequate correction for 
incompleteness effects, as well as for contamination from background
sources.\\ 

With a total exposure time of 201 ks, we detected 263 X-ray point sources
within a distance of 12{\arcmin} from the centre of M31. Of these sources 
64 were not observed previously. This allowed us
to study the sources to a minimum luminosity of $\sim 10^{35}$ erg s$^{-1}$,
whereas the sample is complete above $\sim 10^{36}$ erg s$^{-1}$. We
found good agreement between the observed number of sources and the
expected number, predicted
based on the \textit{K}-band luminosity and average X-ray mass to
light ratio for nearby galaxies. 
The radial distribution of the M31 sources (Figure \ref{fig:spatial})
can be interpreted  as  
superposition of the following three components:  
(i) primordial LMXBs following the \textit{K}-band 
light profile, 
(ii) LMXBs created through dynamical interactions in the inner
bulge of the galaxy, with a distribution that follows the square of the
stellar density $\rho^2_*$ and 
(iii) LMXBs dynamically created in the
globular clusters, with a radial profile that follows the distribution of
globular clusters in M31.
Superimposed on these are the CXB
sources, the distribution of which is flat on the angular scales under
consideration.\\  

\begin{figure}
\resizebox{\hsize}{!}{\includegraphics[angle=0]{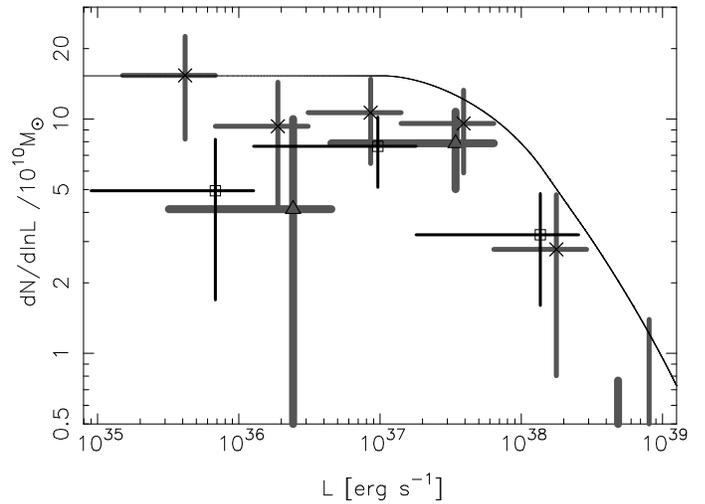}}
\caption{The LF of sources with a radial distance of 
1{\arcmin}-3{\arcmin} (crosses, 28 sources), 3\arcmin-6\arcmin 
(squares, 56 sources) and
6\arcmin-12\arcmin from the centre of M31, excluding sources in
GCs and GC candidates (triangles, 58 sources).
The solid line is the average LMXB LF, with the same normalization
if figures 5-10.}
\label{fig:lumfun_regs}
\end{figure}

\begin{figure}
\resizebox{\hsize}{!}{\includegraphics[angle=0]{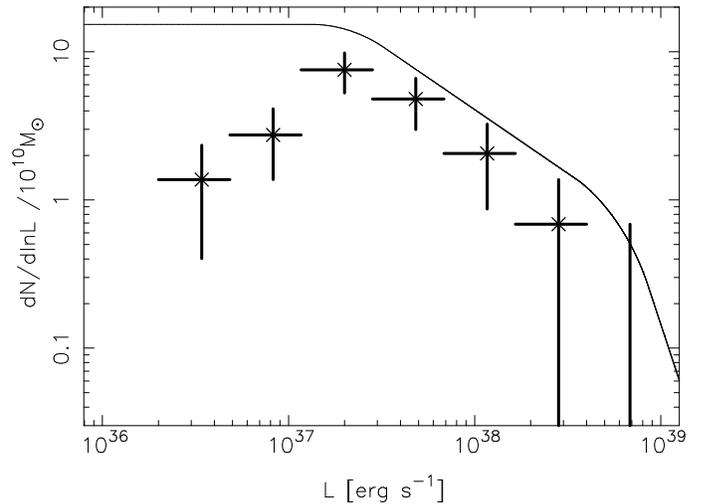}}
\caption{The LF of the maximum luminosity of the transients
observed (28 sources). Note, that below $\sim
10^{37}$ erg/s it is subject to strong selection effects.
The solid line is the average LMXB LF, with the same normalization
in figures 5-10. }
\label{fig:lumfun_trans}
\end{figure}

After applying the incompleteness correction and subtracting the
contribution of CXB sources, we were able to recover the LF of M31
sources down to the luminosity of 
$\sim{\rm few}\times 10^{35}$ erg s$^{-1}$, which is a factor of $\sim$ 3
better than previous studies. 
The luminosity distribution of all X-ray sources in the bulge
of M31 (Fig.\ref{fig:lumfun_all}) is consistent with the average
LMXB LF obtained by \citet{gilfanov}, in particular, it follows the
$dN/dL\propto L^{-1}$ law in the faint luminosity limit, in agreement
with the behaviour found earlier for LMXBs in the Milky Way and in Cen
A. It was furthermore possible to divide
the LMXBs into two subpopulations -- primordial LMXBs and dynamically
formed ones, in order to study the differences in their luminosity
distributions 
(Figure \ref{fig:lumfun_almost}).  
We found that the LF of the primordial LMXBs is consistent with the
average LMXB LF, and is independent of the radial distance from the
centre of M31, within the accuracy allowed by the statistics of our
sample. The LMXBs thought to be of 
dynamical origin have a significantly different luminosity
distribution -- below $\log(L_X)\la 36.5$ their LF shows a prominent
decrease towards low luminosities (Figure \ref{fig:lumfun_dyn}). The
statistics is not sufficient to  
tightly constrain the shape of the LF, in particular to distinguish
between a true low luminosity cut-off and a more moderate flattening
of the luminosity function. It is however sufficient to claim that the
low luminosity, $\log(L_X)<37$, slope of the LF of these sources,
$-0.6\pm0.2$, is inconsistent with the $dN/dL\propto L^{-1}$ law.

We identified the population of transient sources and found that their
radial distribution is consistent with the distribution of persistent
sources. However the current statistics is insufficient to investigate
differences between the fractions of transients in primordial and
dynamically formed LMXBs, especially in the inner 30{\arcsec}, where 
the radial distribution of these two populations differ the most.
Above 
$\sim 10^{37}$  erg s$^{-1}$ the LF of the maximum luminosity of
the transients follows the average LMXB LF (Figure \ref{fig:lumfun_trans}).

\begin{acknowledgements}
This reaserch has made use of \textit{CHANDRA} archival data provided by
the \textit{CHANDRA} X-ray Center and data from the 2MASS Large Galaxy
Atlas provided by NASA/IPAC infrared science archive, as well as
\textit{ASM/RXTE} data obtained through the HEASARC online service.
We thank the referee for helpful remarks on the paper.
\end{acknowledgements}

\begin{appendix}
\section{The effects of source blending on the luminosity function}
\label{sec:blending}
\begin{figure}
\resizebox{\hsize}{!}{\includegraphics[angle=270]{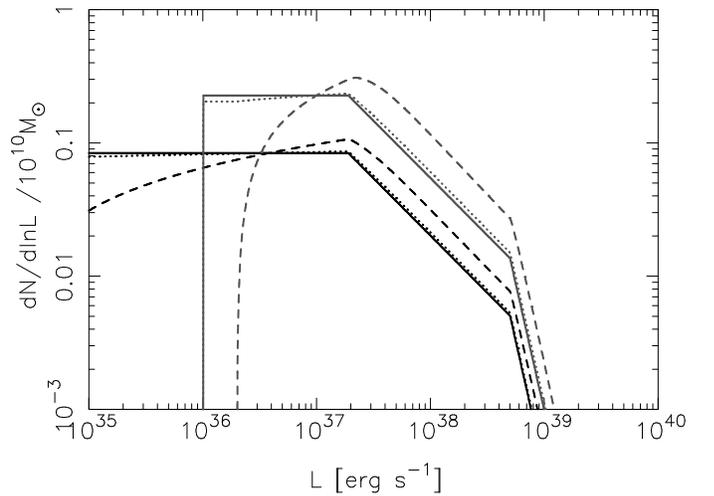}}
\caption{The effects of blending on the LMXB LF of \citet{gilfanov}.
The solid lines are the LFs without blending, whereas the dashed and
dotted lines are the LFs assuming $b=1$ and $b=0.1$, respectively.}
\label{fig:blending}
\end{figure}
In densely populated regions of X-ray point sources, such as GCs or the
very inner parts of a galaxy, a fraction of the observed point sources will
be a blend of two or more sources. Here we present a brief
investigation of the effects of such source blending on the observed LF.
If the distribution of luminosities of single sources is given by $P_1(L)$,
then the distribution of luminosities $P_2(L)$ of a blend of two sources is given
by
\begin{equation}
P_2(L)=\int_0^LP_1(\xi )P_1(L-\xi)d\xi.
\end{equation}
Ignoring blends of three or more sources,
the observed luminosity distribution is then given by
\begin{equation}
\label{eq:blending}
P_d(L)=(1-b)P_1(L)+bP_2(L)
\end{equation}
where $b$ is the fraction of observed sources that are blends of two
sources.
In Figure \ref{fig:blending} we show the effects of source blending on the
LMXB LF of \citet{gilfanov}. The two models shown have different lower cut-off
luminosities, below which the functions are set to zero, $10^{34}$ erg
s$^{-1}$
for the lower model and $10^{36}$ erg s$^{-1}$ for the upper model
(the assumed LFs are therefore equal to the LFs for which $b$ was
calculated in section \ref{sect:LF}). The solid
lines give $P_1(L)$, whereas the dashed lines gives $P_2(L)$. The dotted lines
gives $P_d(L)$, for $b=0.10$. It is clear that for sources samples with
$b\lesssim$0.10 the effects on the observed LF are negligible, and even for samples
with higher values of $b$, the effects are relatively small.

\end{appendix}

{\tiny
\longtab{2}{
\begin{longtable}{ccccrrrccccc}
\caption{The list of point like X-ray sources within a
distance of 12{\arcmin} from the centre of M31}\\
\hline\hline
Number & distance & RA & DEC & cts & corrected cts & error & luminosity & type & id & name \\
(1) & (2) & (3) & (4) & (5) & (6) &  (7) & (8) & (9) & (10) & (11)\\
\hline
\endfirsthead
\caption{continued.}\\
\hline\hline
Number & distance & RA & DEC & cts & corrected cts & error & luminosity & type & id & name \\
(1) & (2) & (3) & (4) & (5) & (6) &  (7) & (8) & (9) & (10) & (11)\\
\hline
\endhead
\hline
\endfoot
\label{sourcelist}
1
 & 1.0 & 00 42 44.37 & 41 16 08.7 & 2245 & 2580.7 & 58.0 & 7.08e+36 &  &  & r1-10\\
2
 & 2.2 & 00 42 44.38 & 41 16 07.4 & 2864 & 3308.2 & 65.4 & 1.08e+37 &  &  & r1-9\\
3
 & 4.0 & 00 42 44.38 & 41 16 05.4 & 1117 & 1233.9 & 41.3 & 5.32e+36 &  &  & r1-21\\
4
 & 4.6 & 00 42 44.30 & 41 16 14.0 & 316 & 344.9 & 22.6 & 1.52e+36 &  &  & r1-22\\
5
 & 5.4 & 00 42 43.86 & 41 16 11.1 & 225 & 244.1 & 19.2 & 1.07e+36 &  &  & r1-27\\
6
 & 7.4 & 00 42 43.87 & 41 16 03.9 & 709 & 811.6 & 33.1 & 3.53e+36 &  &  & r1-23\\
7
 & 9.7 & 00 42 44.68 & 41 16 18.2 & 906 & 1047.3 & 37.3 & 4.62e+36 &  &  & r1-8\\
8
 & 10.6 & 00 42 45.24 & 41 16 11.1 & 456 & 516.2 & 26.8 & 2.25e+36 &  &  & r1-20\\
9
 & 14.5 & 00 42 45.60 & 41 16 08.6 & 1026 & 1191.0 & 39.5 & 5.21e+36 &  &  & r1-7\\
10
 & 15.3 & 00 42 45.12 & 41 16 21.7 & 3098 & 3631.2 & 67.9 & 1.61e+37 & RAD & SIM WSTB 37W135 & r1-4\\
11
 & 20.8 & 00 42 43.88 & 41 16 29.6 & 1293 & 1463.5 & 44.5 & 6.54e+36 &  &  & r1-11\\
12
 & 21.7 & 00 42 46.01 & 41 16 19.6 & 468 & 531.3 & 27.2 & 2.33e+36 &  t &  & r1-19\\
13
 & 23.8 & 00 42 43.75 & 41 16 32.4 & 2329 & 2726.7 & 59.0 & 1.24e+37 &  &  & r1-12\\
14
 & 24.0 & 00 42 42.18 & 41 16 08.3 & 4525 & 5353.9 & 81.6 & 2.45e+37 &  t &  & r1-5\\
15
 & 25.9 & 00 42 42.48 & 41 15 53.7 & 3077 & 3601.1 & 67.6 & 1.62e+37 & PN & CIA 4 & r1-14\\
16
 & 30.1 & 00 42 43.00 & 41 15 43.2 & 2424 & 2839.4 & 60.1 & 1.29e+37 &  &  & r1-13\\
17
 & 30.6 & 00 42 46.97 & 41 16 15.6 & 4037 & 4736.5 & 77.3 & 2.07e+37 &  &  & r1-3\\
18
 & 33.3 & 00 42 43.20 & 41 16 40.3 & 413 & 473.2 & 25.6 & 2.20e+36 &  &  & r1-24\\
19
 & 33.5 & 00 42 46.16 & 41 15 43.2 & 372 & 427.0 & 24.3 & 1.91e+36 &  &  & r1-18\\
20
 & 37.5 & 00 42 47.17 & 41 16 28.4 & 10653 & 12498.5 & 124.6 & 5.59e+37 & PN & CIA 11 & r1-2\\
21
 & 41.0 & 00 42 42.65 & 41 16 45.9 & 38 & 33.6 & 8.8 & 1.58e+35 &  &  & X\\
22
 & 42.4 & 00 42 47.87 & 41 16 22.9 & 370 & 425.5 & 24.2 & 1.92e+36 &  &  & r1-17\\
23
 & 44.8 & 00 42 47.88 & 41 15 49.8 & 264 & 304.3 & 20.7 & 1.40e+36 &  t &  & r1-25\\
24
 & 45.0 & 00 42 42.08 & 41 15 32.0 & 366 & 417.4 & 24.2 & 2.14e+36 &  &  & r1-31\\
25
 & 45.1 & 00 42 45.54 & 41 16 52.3 & 27 & 23.4 & 7.6 & 1.09e+35 &  &  & X\\
26
 & 47.0 & 00 42 45.09 & 41 15 23.2 & 386 & 441.1 & 24.8 & 2.05e+36 & PN & CIA 18 & r1-26\\
27
 & 52.0 & 00 42 48.72 & 41 16 24.5 & 523 & 606.0 & 28.6 & 2.72e+36 &  &  & r1-16\\
28
 & 53.3 & 00 42 40.00 & 41 15 47.5 & 4068 & 4772.3 & 77.5 & 2.32e+37 & RNova & SI 1997-06 & r1-15\\
29
 & 53.5 & 00 42 39.59 & 41 16 14.3 & 2967 & 3500.0 & 66.4 & 1.73e+37 &  t &  & r1-34\\
30
 & 53.9 & 00 42 42.53 & 41 16 59.4 & 210 & 240.8 & 18.6 & 1.11e+36 &  &  & r1-30\\
31
 & 54.5 & 00 42 47.90 & 41 15 32.9 & 3265 & 3834.7 & 69.6 & 1.84e+37 &  &  & r1-6\\
32
 & 55.7 & 00 42 43.79 & 41 15 14.1 & 110 & 117.5 & 13.8 & 5.69e+35 &  t &  & r1-28\\
33
 & 55.8 & 00 42 41.45 & 41 15 23.8 & 471 & 540.7 & 27.2 & 2.93e+36 & GC & Bol B124 & r1-32\\
34
 & 64.7 & 00 42 38.59 & 41 16 03.7 & 27143 & 31287.2 & 198.7 & 1.54e+38 &  &  & r2-26\\
35
 & 67.8 & 00 42 48.53 & 41 15 21.2 & 12231 & 14353.6 & 133.8 & 6.74e+37 &  &  & r1-1\\
36
 & 72.2 & 00 42 50.62 & 41 15 57.1 & 35 & 23.7 & 8.5 & 1.15e+35 & SNR & B90 101 & r2-56\\
37
 & 73.4 & 00 42 39.65 & 41 17 00.7 & 35 & 33.5 & 8.4 & 1.71e+35 &  &  & K\\
38
 & 73.6 & 00 42 45.22 & 41 17 22.3 & 675 & 783.9 & 32.2 & 3.63e+36 &  t &  & r2-16\\
39
 & 75.5 & 00 42 50.82 & 41 15 51.6 & 25 & 22.1 & 7.4 & 1.08e+35 & RNova & B68 27 & X\\
40
 & 75.7 & 00 42 38.80 & 41 15 26.2 & 35 & 29.8 & 8.5 & 1.50e+35 &  &  & r2-54\\
41
 & 76.0 & 00 42 42.73 & 41 14 55.5 & 46 & 47.2 & 9.4 & 2.55e+35 &  t &  & r2-20\\
42
 & 81.9 & 00 42 46.14 & 41 17 28.6 & 21 & 20.7 & 6.8 & 9.45e+34 &  &  & X\\
43
 & 86.8 & 00 42 42.34 & 41 14 45.5 & 1868 & 2198.5 & 53.0 & 1.18e+37 &  &  & r2-21\\
44
 & 89.2 & 00 42 46.09 & 41 17 36.3 & 67 & 74.8 & 11.1 & 3.41e+35 & GCC t & Bol BH16 & r2-15\\
45
 & 90.5 & 00 42 44.91 & 41 17 39.7 & 771 & 901.5 & 34.4 & 4.13e+36 &  &  & r2-18\\
46
 & 97.2 & 00 42 52.53 & 41 15 40.0 & 2740 & 3237.8 & 63.9 & 1.51e+37 &  &  & r2-12\\
47
 & 99.7 & 00 42 52.44 & 41 16 48.7 & 140 & 158.6 & 15.4 & 7.52e+35 &  t &  & X\\
48
 & 107.5 & 00 42 49.15 & 41 17 42.0 & 90 & 103.4 & 12.6 & 4.75e+35 &  &  & r2-41\\
49
 & 113.5 & 00 42 44.63 & 41 18 02.8 & 21 & 19.9 & 6.8 & 9.00e+34 &  &  & X\\
50
 & 114.3 & 00 42 39.54 & 41 14 28.5 & 1169 & 1370.6 & 42.1 & 6.79e+36 &  &  & r2-25\\
51
 & 116.9 & 00 42 34.78 & 41 15 23.3 & 149 & 167.8 & 15.9 & 8.27e+35 &  t &  & r2-28\\
52
 & 117.8 & 00 42 33.90 & 41 16 19.8 & 1217 & 1426.2 & 42.9 & 7.25e+36 &  &  & r2-30\\
53
 & 119.1 & 00 42 39.27 & 41 14 24.7 & 48 & 40.9 & 9.7 & 2.50e+35 &  &  & r2-62\\
54
 & 120.0 & 00 42 54.94 & 41 16 03.2 & 11609 & 13667.2 & 130.1 & 6.21e+37 &  &  & r2-11\\
55
 & 122.6 & 00 42 45.10 & 41 14 07.1 & 618 & 720.8 & 31.0 & 3.48e+36 &  &  & r2-17\\
56
 & 124.3 & 00 42 52.30 & 41 17 35.0 & 122 & 137.6 & 14.5 & 6.37e+35 &  &  & r2-50\\
57
 & 126.2 & 00 42 42.63 & 41 14 04.6 & 24 & 20.3 & 7.3 & 1.04e+35 &  &  & X\\
58
 & 130.6 & 00 42 36.05 & 41 17 41.0 & 15 & 13.0 & 6.0 & 6.04e+34 & GCC & Bol B261 & X\\
59
 & 134.3 & 00 42 33.41 & 41 17 03.5 & 196 & 225.7 & 18.0 & 1.11e+36 &  t &  & r2-70\\
60
 & 134.9 & 00 42 32.53 & 41 15 45.7 & 93 & 104.1 & 12.8 & 4.91e+35 &  &  & r2-55\\
61
 & 138.2 & 00 42 49.24 & 41 18 16.0 & 1627 & 1915.2 & 49.5 & 8.74e+36 &  &  & r2-14\\
62
 & 140.6 & 00 42 50.25 & 41 18 13.1 & 24 & 23.5 & 7.2 & 1.10e+35 &  &  & r2-40\\
63
 & 140.6 & 00 42 40.56 & 41 13 55.3 & 21 & 15.6 & 6.9 & 7.59e+34 &  &  & r2-23\\
64
 & 148.9 & 00 42 31.14 & 41 16 21.7 & 7403 & 8689.1 & 104.2 & 5.21e+37 &  &  & r2-34\\
65
 & 161.5 & 00 42 58.12 & 41 16 52.5 & 21 & 18.6 & 6.9 & 8.64e+34 & GCC & Bol AU010 & X\\
66
 & 162.5 & 00 42 40.22 & 41 18 45.2 & 1090 & 1278.4 & 40.8 & 6.15e+36 &  &  & r2-24\\
67
 & 163.0 & 00 42 58.32 & 41 15 29.2 & 1537 & 1807.0 & 48.2 & 8.39e+36 &  &  & r2-7\\
68
 & 163.9 & 00 42 34.45 & 41 18 09.8 & 94 & 105.3 & 12.9 & 5.01e+35 &  t &  & r2-29\\
69
 & 163.9 & 00 42 36.61 & 41 13 50.3 & 130 & 144.9 & 14.9 & 7.06e+35 &  &  & r2-42\\
70
 & 164.1 & 00 42 30.28 & 41 16 53.2 & 71 & 75.4 & 11.4 & 4.64e+35 &  &  & r2-44\\
71
 & 167.1 & 00 42 40.67 & 41 13 27.4 & 1530 & 1803.7 & 48.1 & 8.75e+36 &  &  & r2-22\\
72
 & 170.2 & 00 42 43.31 & 41 13 19.6 & 844 & 991.6 & 36.0 & 5.02e+36 &  t &  & r2-19\\
73
 & 172.4 & 00 42 52.53 & 41 18 34.8 & 59 & 63.8 & 10.5 & 3.06e+35 &  &  & r2-49\\
74
 & 175.5 & 00 42 59.87 & 41 16 05.8 & 6988 & 8269.1 & 101.4 & 3.76e+37 & GC & Bol B144 & r2-5\\
75
 & 177.1 & 00 42 50.72 & 41 13 27.8 & 20 & 18.4 & 6.7 & 9.03e+34 &  &  & X\\
76
 & 185.8 & 00 42 42.24 & 41 19 13.8 & 37 & 38.7 & 8.6 & 1.83e+35 & RAD & B90 86 & r2-53\\
77
 & 186.3 & 00 42 52.64 & 41 13 28.5 & 60 & 61.4 & 10.6 & 2.98e+35 &  &  & r2-38\\
78
 & 189.2 & 00 42 52.53 & 41 18 54.4 & 13139 & 15495.8 & 138.6 & 7.47e+37 &  &  & r2-13\\
79
 & 191.2 & 00 42 55.19 & 41 18 36.1 & 635 & 744.8 & 31.5 & 3.58e+36 &  &  & r2-10\\
80
 & 192.4 & 00 42 54.36 & 41 13 33.9 & 30 & 28.1 & 7.9 & 1.36e+35 &  &  & X\\
81
 & 193.6 & 00 42 55.62 & 41 18 35.1 & 628 & 735.7 & 31.3 & 3.54e+36 & FGS & Bol B138 & r2-9\\
82
 & 195.7 & 00 42 29.84 & 41 17 57.5 & 19 & 16.7 & 6.6 & 1.01e+35 & GC & Bol B103 & X\\
83
 & 204.1 & 00 42 51.66 & 41 13 02.9 & 260 & 297.7 & 20.6 & 1.51e+36 &  &  & r2-39\\
84
 & 204.5 & 00 42 27.49 & 41 14 53.0 & 76 & 81.3 & 11.7 & 4.54e+35 &  t &  & X\\
85
 & 210.0 & 00 42 56.93 & 41 18 43.9 & 494 & 579.8 & 27.9 & 2.73e+36 &  t &  & r2-8\\
86
 & 211.4 & 00 43 01.78 & 41 17 26.4 & 44 & 46.1 & 9.3 & 2.39e+35 &  &  & r2-46\\
87
 & 211.5 & 00 42 42.82 & 41 19 40.3 & 21 & 19.7 & 6.9 & 9.46e+34 &  &  & X\\
88
 & 215.2 & 00 43 02.94 & 41 15 22.6 & 4541 & 5335.8 & 82.2 & 2.52e+37 & GC & Bol B146 & r2-4\\
89
 & 216.1 & 00 42 26.57 & 41 17 31.5 & 6 & 1.1 & 4.4 & 5.89e+33 &  &  & X\\
90
 & 217.3 & 00 43 03.22 & 41 15 27.8 & 6398 & 7511.3 & 97.2 & 3.57e+37 &  t &  & r2-3\\
91
 & 217.5 & 00 42 58.09 & 41 13 37.3 & 39 & 36.6 & 8.9 & 1.72e+35 &  &  & X\\
92
 & 221.0 & 00 42 32.75 & 41 13 10.9 & 1005 & 1131.2 & 39.3 & 6.11e+36 &  &  & r2-31\\
93
 & 222.8 & 00 42 32.09 & 41 13 14.4 & 5162 & 6067.4 & 87.3 & 3.18e+37 &  &  & r2-32\\
94
 & 222.8 & 00 42 49.03 & 41 19 45.8 & 31 & 28.6 & 8.0 & 1.41e+35 & RAD & SIM WSTB 37W138 & r2-66\\
95
 & 224.9 & 00 43 04.25 & 41 16 01.3 & 969 & 1139.6 & 38.6 & 5.62e+36 & GCC & Fan 42 & r2-1\\
96
 & 229.2 & 00 42 24.19 & 41 15 36.9 & 35 & 26.1 & 8.5 & 1.50e+35 &  &  & r2-52\\
97
 & 229.6 & 00 42 32.09 & 41 19 13.1 & 37 & 35.0 & 8.7 & 1.70e+35 &  &  & X\\
98
 & 230.3 & 00 42 58.10 & 41 13 19.6 & 34 & 31.8 & 8.4 & 1.50e+35 &  &  & X\\
99
 & 232.6 & 00 43 01.71 & 41 18 14.5 & 31 & 26.6 & 8.0 & 1.33e+35 &  &  & r2-47\\
100
 & 234.4 & 00 43 01.12 & 41 13 51.6 & 242 & 270.6 & 19.9 & 1.27e+36 &  &  & r2-37\\
101
 & 235.1 & 00 42 30.96 & 41 19 10.1 & 104 & 117.9 & 13.5 & 6.95e+35 &  &  & r2-43\\
102
 & 240.7 & 00 42 24.24 & 41 17 31.5 & 44 & 38.9 & 9.3 & 2.19e+35 &  &  & r2-57\\
103
 & 243.2 & 00 42 22.96 & 41 15 35.3 & 13877 & 16254.1 & 142.4 & 9.59e+37 &  &  & r3-39\\
104
 & 244.6 & 00 43 01.72 & 41 18 35.5 & 28 & 24.1 & 7.7 & 1.24e+35 &  &  & X\\
105
 & 246.8 & 00 43 05.68 & 41 17 02.7 & 18818 & 21892.3 & 166.0 & 1.05e+38 &  t &  & r2-67\\
106
 & 248.9 & 00 43 03.87 & 41 18 04.9 & 4658 & 5511.9 & 83.1 & 2.81e+37 & GC & Bol B148 & r2-2\\
107
 & 251.0 & 00 42 44.40 & 41 11 58.4 & 488 & 571.2 & 27.7 & 4.36e+36 &  &  & r3-30\\
108
 & 253.7 & 00 42 47.24 & 41 11 57.9 & 159 & 181.2 & 16.4 & 1.34e+36 &  &  & r3-27\\
109
 & 254.3 & 00 42 24.16 & 41 14 15.3 & 22 & 19.9 & 7.0 & 1.21e+35 &  &  & X\\
110
 & 255.9 & 00 42 31.26 & 41 19 38.9 & 2884 & 3395.5 & 65.6 & 2.01e+37 & GC & Bol B107 & r2-33\\
111
 & 256.9 & 00 42 59.66 & 41 19 19.3 & 6646 & 7851.7 & 98.9 & 3.94e+37 & GC & Bol B143 & r2-6\\
112
 & 257.4 & 00 42 21.49 & 41 16 01.3 & 4270 & 5026.6 & 79.5 & 3.20e+37 &  &  & r3-42\\
113
 & 257.6 & 00 42 35.22 & 41 20 05.7 & 1496 & 1753.5 & 47.6 & 9.10e+36 &  &  & r2-27\\
114
 & 259.1 & 00 43 02.92 & 41 18 41.5 & 22 & 14.3 & 7.1 & 7.30e+34 &  &  & X\\
115
 & 260.3 & 00 42 21.29 & 41 15 48.8 & 21 & 15.5 & 6.9 & 1.09e+35 & RAD & B90 65 & X\\
116
 & 262.3 & 00 42 25.15 & 41 13 40.6 & 614 & 712.8 & 31.0 & 4.44e+36 &  &  & r2-45\\
117
 & 262.5 & 00 42 21.80 & 41 15 02.7 & 32 & 29.1 & 8.1 & 1.89e+35 &  &  & X\\
118
 & 266.2 & 00 42 56.04 & 41 12 18.4 & 537 & 634.5 & 29.0 & 4.53e+36 &  t &  & r2-71\\
119
 & 267.5 & 00 42 48.29 & 41 20 33.1 & 24 & 22.6 & 7.2 & 1.11e+35 &  &  & X\\
120
 & 267.9 & 00 42 23.16 & 41 14 07.5 & 1223 & 1435.2 & 43.2 & 8.95e+36 &  &  & r3-38\\
121
 & 269.0 & 00 42 59.52 & 41 12 42.3 & 173 & 191.7 & 17.1 & 9.62e+35 &  &  & r2-48\\
122
 & 270.0 & 00 42 54.79 & 41 20 12.2 & 21 & 17.9 & 6.9 & 8.95e+34 &  &  & X\\
123
 & 271.5 & 00 42 44.85 & 41 11 38.0 & 2955 & 3465.1 & 66.5 & 2.53e+37 &  &  & r3-29\\
124
 & 271.9 & 00 42 57.17 & 41 19 59.5 & 24 & 19.7 & 7.3 & 9.79e+34 &  &  & X\\
125
 & 277.1 & 00 42 26.05 & 41 19 15.0 & 1447 & 1697.3 & 46.9 & 9.40e+36 & GC & Bol B096 & r2-36\\
126
 & 278.9 & 00 42 21.57 & 41 14 19.7 & 684 & 800.4 & 32.6 & 5.30e+36 &  &  & r3-41\\
127
 & 279.8 & 00 42 31.32 & 41 20 07.9 & 111 & 121.3 & 13.9 & 7.27e+35 &  &  & r2-51\\
128
 & 284.0 & 00 43 07.12 & 41 18 10.2 & 166 & 184.7 & 16.8 & 9.91e+35 & Nova t & PIE RJC 99 Jul 98 & r3-115\\
129
 & 285.4 & 00 42 20.85 & 41 17 56.7 & 33 & 31.4 & 8.2 & 1.85e+35 &  &  & X\\
130
 & 287.6 & 00 42 21.09 & 41 18 08.6 & 81 & 87.7 & 12.1 & 5.13e+35 &  t &  & r3-43\\
131
 & 289.5 & 00 42 28.30 & 41 12 23.1 & 6286 & 7379.4 & 96.2 & 5.72e+37 &  &  & r2-35\\
132
 & 291.5 & 00 42 22.44 & 41 13 34.1 & 2253 & 2642.2 & 58.2 & 1.70e+37 &  &  & r3-40\\
133
 & 294.7 & 00 42 33.82 & 41 20 39.3 & 27 & 25.3 & 7.6 & 1.33e+35 &  &  & X\\
134
 & 297.5 & 00 42 58.61 & 41 11 59.5 & 28 & 26.5 & 7.7 & 1.72e+35 &  &  & r2-59\\
135
 & 297.6 & 00 42 41.65 & 41 21 05.5 & 450 & 522.2 & 26.7 & 3.64e+36 &  &  & r3-31\\
136
 & 298.2 & 00 42 47.83 & 41 11 13.9 & 168 & 189.4 & 16.8 & 1.50e+36 & GCC t & Bol B128 & X\\
137
 & 300.6 & 00 42 59.02 & 41 11 58.8 & 33 & 32.8 & 8.2 & 2.14e+35 &  &  & r2-58\\
138
 & 301.6 & 00 42 20.85 & 41 13 44.5 & 27 & 23.2 & 7.6 & 1.62e+35 &  &  & X\\
139
 & 306.8 & 00 43 10.62 & 41 14 51.4 & 14308 & 16890.8 & 144.8 & 8.99e+37 & GC & Bol B153 & r3-15\\
140
 & 307.1 & 00 42 50.02 & 41 11 09.1 & 93 & 98.7 & 12.9 & 7.51e+35 &  &  & r3-24\\
141
 & 308.9 & 00 42 41.12 & 41 11 02.6 & 30 & 25.9 & 8.0 & 2.07e+35 &  &  & r3-32\\
142
 & 311.2 & 00 42 46.94 & 41 21 19.2 & 214 & 243.8 & 18.8 & 1.71e+36 &  &  & r3-28\\
143
 & 311.9 & 00 42 18.43 & 41 17 59.6 & 8 & 0.8 & 4.9 & 5.37e+33 &  &  & X\\
144
 & 312.2 & 00 43 06.80 & 41 19 11.6 & 89 & 83.8 & 12.7 & 3.98e+35 & EXT & Source 231 & r3-67\\
145
 & 312.9 & 00 42 16.55 & 41 16 10.7 & 30 & 27.5 & 7.9 & 1.90e+35 &  &  & r3-75\\
146
 & 313.5 & 00 42 17.04 & 41 15 08.2 & 261 & 299.0 & 20.6 & 2.19e+36 &  t &  & r3-46\\
147
 & 314.6 & 00 43 06.75 & 41 19 16.6 & 100 & 97.3 & 13.3 & 4.73e+35 & EXT & Source 160 & r3-67\\
148
 & 316.2 & 00 42 18.65 & 41 14 01.9 & 5388 & 6324.4 & 89.2 & 4.69e+37 & GC & Bol B086 & r3-44\\
149
 & 318.6 & 00 42 16.09 & 41 15 53.3 & 37 & 36.6 & 8.6 & 2.80e+35 &  &  & r3-76\\
150
 & 318.9 & 00 42 47.88 & 41 10 53.1 & 54 & 56.4 & 10.1 & 4.71e+35 &  &  & r3-26\\
151
 & 320.1 & 00 43 02.44 & 41 12 03.1 & 54 & 55.6 & 10.1 & 4.05e+35 &  &  & r3-68\\
152
 & 321.4 & 00 42 20.48 & 41 13 13.2 & 34 & 27.6 & 8.4 & 2.03e+35 &  &  & r3-89\\
153
 & 327.9 & 00 43 11.37 & 41 18 09.9 & 317 & 360.6 & 22.7 & 1.93e+36 &  &  & r3-14\\
154
 & 330.4 & 00 42 15.69 & 41 17 21.0 & 2105 & 2476.8 & 56.2 & 1.69e+37 &  &  & r3-47\\
155
 & 335.1 & 00 43 09.85 & 41 19 00.9 & 1328 & 1559.0 & 45.0 & 8.29e+36 & GCC t & Fan 44 & r3-16\\
156
 & 335.6 & 00 42 27.71 & 41 20 48.1 & 68 & 68.7 & 11.2 & 4.54e+35 &  &  & r3-37\\
157
 & 338.5 & 00 42 40.68 & 41 10 33.4 & 154 & 167.0 & 16.2 & 1.35e+36 & GC & Bol B123 & r3-34\\
158
 & 339.3 & 00 43 13.88 & 41 17 12.2 & 70 & 64.0 & 11.4 & 3.59e+35 &  &  & r3-12\\
159
 & 340.2 & 00 43 08.64 & 41 12 48.4 & 1109 & 1292.5 & 41.2 & 6.86e+36 &  &  & r3-17\\
160
 & 341.1 & 00 42 57.91 & 41 11 04.8 & 4128 & 4861.9 & 78.4 & 3.65e+37 &  &  & r3-22\\
161
 & 341.5 & 00 43 14.38 & 41 16 50.2 & 119 & 121.6 & 14.5 & 6.61e+35 &  &  & r3-11\\
162
 & 343.1 & 00 42 50.76 & 41 10 34.1 & 48 & 46.0 & 9.7 & 3.58e+35 & GCC & Bol BH18 & r3-71\\
163
 & 344.5 & 00 43 03.02 & 41 20 42.0 & 302 & 344.7 & 22.1 & 1.96e+36 & PN & CIA 165 & r3-21\\
164
 & 344.7 & 00 42 52.26 & 41 21 42.3 & 27 & 23.7 & 7.6 & 1.75e+35 &  &  & X\\
165
 & 346.2 & 00 42 15.24 & 41 18 01.3 & 190 & 210.4 & 17.8 & 1.48e+36 &  &  & r3-49\\
166
 & 348.8 & 00 43 13.23 & 41 18 13.5 & 418 & 473.5 & 25.8 & 2.67e+36 & RNova & CFN 26 & r3-13\\
167
 & 350.4 & 00 42 16.98 & 41 18 56.4 & 54 & 51.6 & 10.2 & 3.24e+35 &  &  & r3-91\\
168
 & 350.6 & 00 42 33.46 & 41 21 38.0 & 49 & 45.2 & 9.8 & 3.83e+35 & FGS & Bol B113 & X\\
169
 & 352.7 & 00 42 13.07 & 41 16 27.9 & 79 & 80.3 & 12.0 & 5.72e+35 &  &  & r3-53\\
170
 & 358.9 & 00 42 34.17 & 41 21 49.7 & 267 & 302.6 & 20.9 & 2.38e+36 &  &  & r3-35\\
171
 & 361.8 & 00 43 16.35 & 41 16 30.5 & 30 & 20.4 & 8.0 & 1.39e+35 &  &  & X\\
172
 & 362.0 & 00 43 07.50 & 41 20 19.9 & 550 & 632.5 & 29.4 & 3.51e+36 & GCC & Bol B150 & r3-18\\
173
 & 366.3 & 00 42 40.73 & 41 10 05.3 & 33 & 25.8 & 8.3 & 2.24e+35 &  &  & X\\
174
 & 366.5 & 00 42 11.99 & 41 16 48.7 & 402 & 458.1 & 25.3 & 3.31e+36 &  &  & r3-55\\
175
 & 368.7 & 00 42 40.92 & 41 22 16.1 & 52 & 48.9 & 10.0 & 3.51e+35 &  &  & r3-33\\
176
 & 369.4 & 00 42 19.01 & 41 20 04.3 & 63 & 58.3 & 10.9 & 3.53e+35 &  &  & r3-90\\
177
 & 369.5 & 00 42 18.37 & 41 12 23.8 & 3119 & 3651.3 & 68.2 & 3.50e+37 &  &  & r3-45\\
178
 & 372.3 & 00 43 01.70 & 41 10 52.9 & 48 & 43.8 & 9.7 & 3.55e+35 &  &  & r3-96\\
179
 & 378.3 & 00 42 12.18 & 41 17 58.7 & 1554 & 1800.7 & 48.5 & 1.20e+37 & GCC & Bol B078 & r3-54\\
180
 & 378.6 & 00 43 03.31 & 41 21 21.7 & 1486 & 1745.3 & 47.5 & 1.34e+37 & FGS & Bol B147 & r3-19\\
181
 & 380.7 & 00 42 13.16 & 41 18 36.5 & 6027 & 7031.2 & 94.3 & 4.91e+37 & FGS & Cra 13 & r3-52\\
182
 & 383.1 & 00 43 15.06 & 41 13 26.6 & 44 & 31.8 & 9.4 & 1.82e+35 &  &  & X\\
183
 & 385.5 & 00 43 14.18 & 41 13 02.0 & 51 & 35.7 & 10.0 & 2.04e+35 &  &  & X\\
184
 & 388.1 & 00 42 10.29 & 41 15 09.9 & 555 & 640.8 & 29.5 & 4.99e+36 &  &  & r3-58\\
185
 & 389.2 & 00 43 16.10 & 41 18 41.3 & 411 & 469.3 & 25.6 & 3.35e+36 &  &  & r3-9\\
186
 & 390.0 & 00 42 49.36 & 41 22 35.2 & 45 & 38.4 & 9.4 & 2.99e+35 &  &  & X\\
187
 & 392.9 & 00 42 15.14 & 41 12 34.6 & 2746 & 3194.3 & 64.2 & 2.67e+37 &  &  & r3-50\\
188
 & 393.2 & 00 42 25.35 & 41 10 39.4 & 33 & 16.3 & 8.4 & 1.59e+35 &  &  & X\\
189
 & 396.2 & 00 43 14.60 & 41 19 30.5 & 61 & 49.2 & 10.8 & 2.82e+35 &  &  & X\\
190
 & 403.9 & 00 42 09.51 & 41 17 45.6 & 1077 & 1244.1 & 40.6 & 8.92e+36 & GCC & Mita 140 & r3-59\\
191
 & 404.3 & 00 42 54.26 & 41 09 41.0 & 29 & 22.5 & 7.9 & 1.91e+35 &  &  & X\\
192
 & 411.3 & 00 42 28.22 & 41 10 00.4 & 1989 & 2316.9 & 54.8 & 2.45e+37 &  &  & r3-36\\
193
 & 411.5 & 00 42 12.69 & 41 12 44.1 & 62 & 51.9 & 10.9 & 4.14e+35 &  &  & r3-92\\
194
 & 411.8 & 00 43 12.35 & 41 20 33.5 & 51 & 31.7 & 10.1 & 1.85e+35 &  &  & X\\
195
 & 412.7 & 00 43 03.13 & 41 10 15.6 & 561 & 643.6 & 29.7 & 5.25e+36 & GCC & Fan 37 & r3-20\\
196
 & 417.2 & 00 42 15.51 & 41 20 31.5 & 570 & 655.6 & 29.9 & 4.67e+36 & GCC & Fan 16 & r3-48\\
197
 & 420.2 & 00 43 21.57 & 41 15 57.4 & 52 & 37.0 & 10.1 & 2.37e+35 &  &  & r3-104\\
198
 & 425.4 & 00 42 07.10 & 41 17 20.1 & 140 & 145.1 & 15.5 & 1.10e+36 &  &  & r3-79\\
199
 & 426.1 & 00 42 11.00 & 41 12 48.3 & 270 & 296.3 & 21.0 & 2.34e+36 &  &  & r3-57\\
200
 & 426.6 & 00 43 21.08 & 41 17 50.5 & 724 & 826.5 & 33.6 & 5.02e+36 &  &  & r3-7\\
201
 & 430.6 & 00 42 07.77 & 41 18 14.9 & 4275 & 4995.6 & 79.7 & 3.72e+37 &  &  & r3-61\\
202
 & 432.2 & 00 43 17.06 & 41 12 25.1 & 28 & 19.9 & 7.8 & 1.75e+35 &  &  & X\\
203
 & 438.5 & 00 42 32.04 & 41 23 05.5 & 68 & 66.5 & 11.3 & 5.82e+35 &  &  & r3-86\\
204
 & 441.7 & 00 42 19.73 & 41 21 53.5 & 93 & 90.4 & 13.0 & 8.51e+35 & GCC & Mita 166 & r3-74\\
205
 & 443.4 & 00 43 20.92 & 41 18 51.7 & 97 & 87.1 & 13.2 & 5.20e+35 &  &  & r3-66\\
206
 & 448.6 & 00 42 59.00 & 41 09 12.6 & 43 & 32.3 & 9.3 & 2.76e+35 &  &  & X\\
207
 & 448.9 & 00 43 19.98 & 41 12 50.2 & 50 & 32.3 & 10. & 2.18e+35 &  &  & X\\
208
 & 451.6 & 00 43 15.42 & 41 11 25.3 & 167 & 182.8 & 16.8 & 1.57e+36 & GC & Bol B161 & r3-10\\
209
 & 453.6 & 00 42 08.21 & 41 12 49.6 & 90 & 82.6 & 12.8 & 6.90e+35 &  &  & r3-93\\
210
 & 453.8 & 00 42 15.06 & 41 21 21.4 & 55 & 46.2 & 10.3 & 5.13e+35 & GCC & Fan 15 & r3-51\\
211
 & 454.4 & 00 42 04.14 & 41 15 32.6 & 164 & 174.2 & 16.7 & 1.42e+36 &  &  & r3-62\\
212
 & 457.2 & 00 42 33.90 & 41 23 31.3 & 160 & 168.9 & 16.5 & 1.34e+36 &  t &  & X\\
213
 & 458.9 & 00 42 09.62 & 41 20 09.8 & 76 & 68.7 & 11.9 & 5.43e+35 & RAD & B90 34 & r3-102\\
214
 & 461.6 & 00 43 18.89 & 41 20 17.0 & 170 & 169.1 & 17.0 & 1.07e+36 &  &  & r3-8\\
215
 & 463.3 & 00 43 24.82 & 41 17 27.3 & 308 & 335.5 & 22.4 & 2.11e+36 &  &  & r3-6\\
216
 & 463.5 & 00 42 05.72 & 41 13 29.9 & 64 & 45.6 & 11.1 & 3.86e+35 &  t &  & r3-125\\
217
 & 464.3 & 00 43 19.54 & 41 20 10.1 & 94 & 80.2 & 13.1 & 4.93e+35 &  &  & X\\
218
 & 468.0 & 00 43 06.65 & 41 22 43.9 & 87 & 87.3 & 12.6 & 7.83e+35 & EmO & MLA 686 & r3-83\\
219
 & 468.4 & 00 43 18.39 & 41 11 41.8 & 38 & 29.3 & 8.8 & 2.69e+35 &  &  & X\\
220
 & 469.6 & 00 43 22.34 & 41 12 58.3 & 61 & 42.5 & 10.8 & 2.76e+35 &  &  & r3-82\\
221
 & 479.9 & 00 43 00.24 & 41 08 44.4 & 48 & 41.6 & 9.7 & 4.33e+35 &  &  & X\\
222
 & 480.5 & 00 43 17.95 & 41 11 14.7 & 42 & 33.0 & 9.2 & 3.33e+35 &  &  & X\\
223
 & 480.7 & 00 42 48.95 & 41 24 07.3 & 109 & 106.5 & 13.9 & 6.95e+35 &  &  & r3-84\\
224
 & 482.1 & 00 43 24.13 & 41 13 14.3 & 79 & 60.5 & 12.1 & 4.02e+35 &  &  & r3-65\\
225
 & 482.6 & 00 42 16.13 & 41 22 12.9 & 41 & 28.6 & 9.1 & 3.88e+35 &  &  & X\\
226
 & 484.8 & 00 42 09.10 & 41 20 48.0 & 698 & 799.3 & 33.0 & 7.81e+36 &  &  & r3-60\\
227
 & 487.2 & 00 42 11.78 & 41 10 49.0 & 386 & 430.9 & 24.9 & 6.10e+36 &  &  & r3-56\\
228
 & 493.0 & 00 43 18.73 & 41 21 13.8 & 48 & 41.3 & 9.7 & 4.07e+35 &  &  & X\\
229
 & 504.0 & 00 42 45.80 & 41 24 33.1 & 96 & 91.9 & 13.1 & 6.76e+35 &  &  & r3-72\\
230
 & 507.5 & 00 43 26.34 & 41 19 11.6 & 151 & 152.5 & 16.1 & 1.25e+36 & RAD & GLG 005 & r3-64\\
231
 & 511.2 & 00 43 27.92 & 41 18 29.9 & 583 & 654.2 & 30.3 & 5.39e+36 & SNR & B90 142 & r3-63\\
232
 & 513.6 & 00 42 05.90 & 41 11 33.7 & 85 & 75.6 & 12.5 & 1.10e+36 &  &  & r3-80\\
233
 & 521.7 & 00 42 24.51 & 41 24 01.1 & 58 & 43.6 & 10.6 & 4.12e+35 &  &  & X\\
234
 & 529.2 & 00 43 10.00 & 41 23 32.5 & 62 & 51.3 & 10.8 & 5.89e+35 & PN t & CIA 350 & r3-127\\
235
 & 537.6 & 00 42 07.58 & 41 10 27.1 & 104 & 93.0 & 13.7 & 1.45e+36 &  &  & r3-94\\
236
 & 548.6 & 00 42 40.98 & 41 07 02.0 & 59 & 49.5 & 10.6 & 7.54e+35 &  &  & r3-85\\
237
 & 554.4 & 00 42 48.50 & 41 25 21.8 & 4302 & 5020.8 & 80.1 & 3.91e+37 &  t &  & r3-25\\
238
 & 565.0 & 00 42 23.00 & 41 07 38.2 & 135 & 122.1 & 15.4 & 2.06e+36 &  &  & r3-73\\
239
 & 571.6 & 00 43 33.83 & 41 14 07.3 & 70 & 46.2 & 11.5 & 4.29e+35 &  &  & X\\
240
 & 575.9 & 00 42 49.41 & 41 06 36.3 & 54 & 42.5 & 10.2 & 7.28e+35 &  &  & r3-98\\
241
 & 585.2 & 00 43 08.89 & 41 07 34.2 & 38 & 34.9 & 8.8 & 8.72e+35 &  &  & X\\
242
 & 587.9 & 00 43 34.32 & 41 13 23.5 & 2185 & 2554.7 & 57.4 & 2.41e+37 &  &  & r3-2\\
243
 & 589.9 & 00 43 35.92 & 41 14 32.8 & 446 & 431.6 & 27.0 & 2.54e+36 &  &  & X\\
244
 & 590.7 & 00 42 53.68 & 41 25 50.6 & 313 & 322.0 & 22.6 & 3.66e+36 & SNR & MG BA521 & r3-69\\
245
 & 600.3 & 00 42 55.40 & 41 25 56.5 & 1483 & 1696.3 & 47.6 & 1.84e+37 &  &  & r3-23\\
246
 & 600.7 & 00 41 51.65 & 41 14 38.7 & 385 & 426.6 & 24.9 & 5.60e+36 &  &  & r3-81\\
247
 & 603.5 & 00 43 37.28 & 41 14 43.4 & 5112 & 5992.7 & 87.0 & 5.65e+37 & GC & Bol B185 & r3-1\\
248
 & 617.2 & 00 42 26.26 & 41 25 52.2 & 581 & 660.6 & 30.2 & 9.26e+36 & EmO & W2 & r3-87\\
249
 & 627.5 & 00 41 50.35 & 41 13 36.3 & 209 & 208.4 & 18.7 & 3.21e+36 &  &  & r3-110\\
250
 & 627.5 & 00 43 14.38 & 41 07 21.6 & 1405 & 1650.1 & 46.2 & 6.12e+37 & GC t & Bol B158 & r3-112\\
251
 & 630.4 & 00 42 51.60 & 41 26 34.5 & 95 & 100.0 & 13.0 & 1.47e+36 & RAD & GLG 011 & r3-70\\
252
 & 634.1 & 00 43 32.38 & 41 10 41.0 & 1298 & 1503.3 & 44.5 & 3.12e+37 &  &  & r3-3\\
253
 & 642.2 & 00 43 14.59 & 41 25 13.5 & 94 & 79.2 & 13.1 & 8.89e+35 & GC & Bol B159 & r3-105\\
254
 & 645.6 & 00 43 33.58 & 41 21 39.1 & 70 & 73.3 & 11.4 & 1.38e+36 &  &  & X\\
255
 & 646.9 & 00 42 37.97 & 41 05 26.5 & 60 & 36.4 & 10.8 & 1.32e+36 &  &  & r3-100\\
256
 & 660.4 & 00 43 41.75 & 41 14 00.9 & 79 & 50.7 & 12.2 & 5.81e+35 &  &  & X\\
257
 & 669.7 & 00 42 32.82 & 41 27 06.5 & 26 & 18.0 & 7.5 & 3.02e+35 &  &  & X\\
258
 & 676.4 & 00 42 10.90 & 41 06 47.8 & 151 & 103.2 & 16.4 & 1.95e+36 &  &  & r3-78\\
259
 & 685.1 & 00 42 20.57 & 41 26 40.2 & 170 & 182.6 & 17.0 & 3.49e+36 & EmO & W2 & K\\
260
 & 691.1 & 00 41 55.15 & 41 23 02.8 & 87 & 57.2 & 12.8 & 2.40e+36 &  &  & r3-108\\
261
 & 706.9 & 00 43 43.94 & 41 12 31.9 & 120 & 108.7 & 14.6 & 1.70e+36 & RAD & B90 166 & X\\
262
 & 715.3 & 00 42 28.98 & 41 04 35.4 & 785 & 807.6 & 35.1 & 2.45e+37 &  &  & r3-111\\
263
 & 718.2 & 00 42 44.29 & 41 28 07.6 & 84 & 71.5 & 12.4 & 1.17e+36 &  &  & X\\
\hline
\end{longtable}
(1) -- The sequence number;
(2) -- Distance to the centre in arcsec;
(3),(4) -- Right ascension and declination of source;
(5) -- Source counts;
(6) -- Source counts after background subtraction;
(7) -- Statistical error on source counts after background subtraction;
(8) -- X-ray luminosity, 0.5-8 keV, assuming 780 kpc distance; 
(9) -- Source Type: GC -- confirmed globular cluster, GCC -- globular cluster
candidate, PN -- planetary nebula, FGS -- foreground star, Nova -- nova,
Rnova -- Nova (random match),
EmO -- emission line object, RAD -- radio source, SNR -- supernova remnant,
EXT -- extended source, t -- transient source;
(10) -- precise identification and reference: Bol -- \citet{bologna}, Fan
-- \citet{fan}, Mita -- \citet{magnier}, MLA -- \citet{meyssonnier}, W2 --
\citet{williams2}, CIA -- \citet{ciardullo}, CFN -- \citet{ciardullo2}, PIE -- \citet{pietsch},
SI -- \citet{shafter}, B68 -- \citet{borngen}, SIM -- Simbad, GLG --
\citet{gelfand}, B90 -- \citet{braun}, MG -- \citet{magnier_snr}, Cra -- \citet{crampton};
(11) -- Source name in \citet{kong}, \citet{williams} and \citet{williams3};
Sources not included in these catalogues are marked with K if observed in
\citet{kaaret}, else with X, indicating that these are new sources.
}
}

\end{document}